\documentclass[10pt,reqno,oneside]{amsart}
\usepackage{verbatim,amsmath,amssymb,cite,epsfig,color,hyperref}
\usepackage[T1]{fontenc}

\usepackage{setspace}
\numberwithin{equation}{section}

\newcommand{\Real}{\mathbb R}
\newcommand{\N}{\mathbb N}

\newtheorem{lemma}{Lemma}[section]
\newtheorem{remark}{Remark}[section]

\begin{document}

\title[Fractal powers in Serrin's vortices]{Fractal powers in Serrin's swirling vortex solutions}

\author{Pavel B\v{e}l\'{\i}k}
\author{Douglas P.~Dokken}
\author{Kurt Scholz}
\author{Mikhail M.~Shvartsman}

\address{Pavel B\v{e}l\'{\i}k\\
Mathematics Department\\
Augsburg College\\
2211 Riverside Avenue\\
Minneapolis, MN 55454\\
U.S.A.} \email{belik@augsburg.edu}

\address{Douglas P.~Dokken\\
Department of Mathematics\\
University of St.~Thomas\\
2115 Summit Avenue\\
St.~Paul, MN 55105\\
U.S.A.} \email{dpdokken@stthomas.edu}

\address{Kurt Scholz\\
Department of Mathematics\\
University of St.~Thomas\\
2115 Summit Avenue\\
St.~Paul, MN 55105\\
U.S.A.} \email{k9scholz@stthomas.edu}

\address{Mikhail M.~Shvartsman\\
Department of Mathematics\\
University of St.~Thomas\\
2115 Summit Avenue\\
St. Paul, MN 55105\\
U.S.A.} \email{mmshvartsman@stthomas.edu}

\thanks{
}



\date{\today}

\begin{abstract}
We consider a modification of the fluid flow model for a tornado-like swirling vortex developed by J.~Serrin \cite{serrin}, where velocity decreases as the reciprocal of the distance from the vortex axis. Recent studies, based on radar data of selected severe weather events \cite{cai,wurman00,wurman05}, indicate that the angular momentum in a tornado may not be constant with the radius, and thus suggest a different scaling of the velocity/radial distance dependence.

Motivated by this suggestion, we consider Serrin's approach with the assumption that the velocity decreases as the reciprocal of the distance from the vortex axis to the power $b$ with a general $b>0$. This leads to a boundary-value problem for a system of nonlinear differential equations. We analyze this problem for particular cases, both with nonzero and zero viscosity, discuss the question of existence of solutions, and use numerical techniques to describe those solutions that we cannot obtain analytically.
\end{abstract}

\maketitle{\allowdisplaybreaks\thispagestyle{empty}

\section{Introduction}
\label{sec:introduction}
Rotating thunderstorms, also known as supercells, and tornadoes generated from them have been modeled using axisymmetric flows. A variety of approaches to investigate axisymmetric flows has led to various models of vortex dynamics \cite{newton01,shapiromarkowski99}. Among the most prominent ones are Rankine combined, Burgers--Rott, Lamb--Oseen, and Sullivan vortex models. Some of these models (e.g., Burgers--Rott) balance vorticity diffusion and advection mechanisms that are important to modeling the inner core of tornadoes and other intense vortices. Most of the models describe rotation in the whole space and therefore they do not take into account friction resulting from contact with the ground. See \cite{shapiromarkowski99} for a detailed list of various axisymmetric models, some of which are exact solutions to Navier--Stokes equations.

In $1972$, J.~Serrin, following the works of Long \cite{long58,long61} and Go\v{l}dshtik \cite{goldshtik60}, discovered a special class of tornado-like swirling vortex solutions to the Navier--Stokes equations in half-space \cite{serrin}, in which the velocity decreases as the reciprocal of the radial distance, $r$, from the vortex axis, a phenomenon observed in real tornadoes \cite{wurman96,shapiromarkowski99}. Serrin's solutions, unlike Long's, model the interaction of a swirling vortex with the horizontal boundary, and they are some of the few exact solutions of Navier--Stokes equations in half-space, in which both the impermeability and the no-slip condition are enforced on a rigid horizontal boundary representing the ground. This should be contrasted with, for example, the popular Burgers--Rott or Sullivan models, in which the no-slip condition is violated. Serrin described three types of solutions depending on the values of kinematic viscosity and a ``pressure'' parameter: downdraft core with radial outflow, updraft core with radial inflow (single-cell vortices), and downdraft core with a compensating radial inflow (double-cell vortex). See Fig.~\ref{fig:breakdown} for a sketch of a single-cell and a double-cell vortex. While these solutions may not be accurate near the vortex core due to the singularity along the vortex axis, outside the region of the most intense winds they seem to provide a reasonable description of a tornado \cite{serrin,shapiromarkowski99,wurman96}. In fact, in \cite{lewellens07}, the authors note that their solution was similar to a similarity solution of Long \cite{long58,long61}, and Serrin's computations are analogous to Long's. Also, as stated in \cite{shapiromarkowski99}, ``The near-surface flow of Serrin's vortex beyond the core region may be a useful analog for the frictional boundary layer in the region of tornadoes beyond the radius of maximum wind [speed].'' Regarding the inner core, the singularity near the vortex axis present in Serrin's model is not present in the Burgers--Rott, Sullivan, or Long's models. On the other hand, some numerical, radar, and ground velocity tracking studies suggest that updraft wind speeds near the tornado axis can achieve large values, approaching and possibly exceeding the speed of sound \cite{bluestein07,fiedler94,fiedler96,fiedlergarfield10,fiedlerrotunno86,lewellens07a,lewellenxialewellen02,xialewellens03,wurman13}.
\begin{figure}
  \centering
  \includegraphics[width=0.9\textwidth]{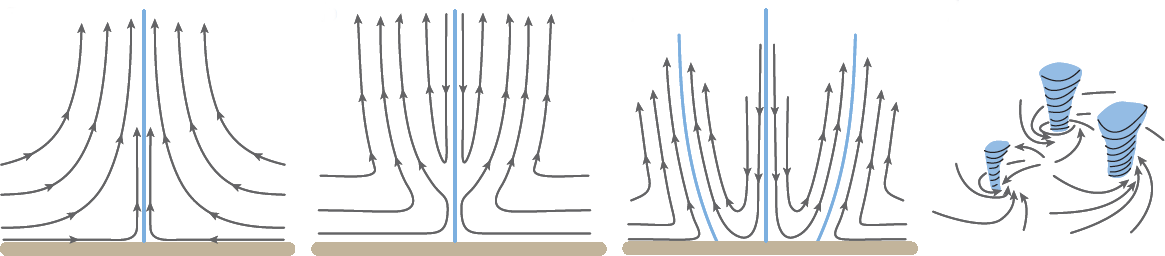}
  \caption{An illustration of a vortex breakdown process. Viewed from left to right, the flow undergoes several bifurcations from a single-cell vortex on the left to multiple vortices on the right. The middle two images show a single-cell vortex below and a double-cell vortex above (left) and a double-cell vortex (right). Modified with permission from \cite{rotunno13}.}
  \label{fig:breakdown}
\end{figure}

The search for axisymmetric flow solutions has continued through the last few decades because of their importance in modeling a wide range of phenomena. Particular types of tornado-like conical solutions influenced by Serrin's work can be found in \cite{wu86,yih82,goldshtikshtern88}. Relevant reviews are presented in \cite{shternhussain99,wang91}. These solutions all exhibit velocity decay reciprocal to the distance from the vortex axis.

However, more recent high-resolution mobile Doppler radar studies \cite{wurman00,wurman05} provide evidence that the velocities decay as the reciprocal of a different power of the radial distance from the vortex axis.
In these papers, devoted to analyzing data associated to strong or violent tornadoes, an attempt is made to calculate the value of the exponent in the ``velocity power law'' $v\propto r^b$. Wurman and Alexander \cite{wurman05} calculated the exponent $b$ from radar data obtained in an intercept of the May 31, 1998 Spencer, South Dakota, tornado and obtained the value $b=-0.67$ for the velocity field away from the core-flow region, in which data indicated a solid-body rotation. The tornado was rated EF4. These values were calculated from the data taken at one instant during the tornado's existence. In the case of the June 2, 1995 Dimmit, Texas, tornado, Wurman and Gill \cite{wurman00} observed exponent values in the range $-0.5$ to $-0.7$, concluding that ``it implies that the angular momentum in the tornado was not constant with radius, but decays toward the center.''

In an attempt to distinguish between tornadic and non-tornadic storms, Cai observed \cite{cai} that tornado-related vorticity fields might have a fractal nature with respect to the grid size; more specifically, natural log of the vorticity, $\zeta$, and natural log of the grid spacing seem to have a linear relationship, with a constant negative ratio. Larger absolute values of this ratio correspond to stronger storms. In some cases of tornadic storms, the ratio is found close to $-1.6$. For tornadic mesocyclones, this suggests a power law of the form $\zeta\propto r^\beta$. Since $\zeta=\nabla\times{v}$, the results of Wurman et al.~and Cai appear to be consistent, even though they consider different scales. We further explore the potential scale invariance between the mesocyclone and tornado scales in a related work \cite{dokken12}. Cai also noted that the exponent in the power law for vorticity can be thought of as measuring a fractal dimension associated with the vortex. The possibility of fractalization of a vortex undergoing stretching was pointed out by Chorin \cite{chorin}. In \cite{dokken12}, we also explore the relationship between vortex stretching and a vortex breakdown. Additional discussions of a vortex breakdown can be found in \cite{bluestein07,fiedler94,fiedlerrotunno86,lewellens07,lewellens07a,lewellensxia00,xialewellens03}. We briefly comment on how our results relate to a vortex breakdown (as illustrated in Fig.~\ref{fig:breakdown}) in the conclusions section.

We therefore find it natural to ask whether there are Serrin-type similarity solutions to the Navier--Stokes equations of the form described in \eqref{eq:velFGO}, in which the velocity field is proportional to $r^{-b}$, where $b>0$ and $b\neq1$, with the most interesting case being $0<b<1$. This work attempts to answer this question. Although other models could conceivably be used as well to try to derive a first model with a velocity decay different from $r^{-1}$, we use Serrin's model as a starting point for its mathematical simplicity and for being an exact solution to the Navier--Stokes equations satisfying the boundary conditions at the ground. We show that under the assumption of constant nonzero viscosity and suitable assumptions on the form of the velocity field, similar to that in \cite{serrin}, the Navier--Stokes equations do not admit any nontrivial solutions except when $b=1$, the case studied by Serrin. However, in the relaxed case of zero viscosity, the Euler equations always admit a simple, purely rotational solution with azimuthal velocity of the form $Cr^{-b}$. In addition, when $b=1$, another set of nontrivial solutions is found analytically. When $b\ge2$, we show that no other solutions exist. The most intriguing cases are when $0<b<1$ and $1<b<2$, for which we have not found analytic solutions; for the former case we present numerically computed solutions for various values of the parameter $b$,
while for the latter case we show that any solution would have to be unstable in the sense of Rayleigh's circulation criterion \cite{drazinreid}. We summarize the main results below.

\begin{table}[h]
  \caption{Summary of the main results for various values of $b$ and viscosity $\nu$}
  \begin{tabular}{|c|c|c|}
    \hline
    $b$      & $\nu>0$ (Sections \ref{sec:NSb=2}, \ref{sec:NSbne12}) & $\nu=0$ \\
    \hline\hline
    $0<b<1$  & no solutions                                          & Section \ref{sec:b<1}: solutions approximated numerically \\
    $b=1$    & Serrin \cite{serrin}                                  & Section \ref{sec:b=1}: all analytic solutions determined \\
    $1<b<2$  & no solutions                                          & Section \ref{sec:b>1}: all solutions must be unstable \\
    $2\le b$ & no solutions                                          & Section \ref{sec:b=2} and \ref{sec:b>2}: no nontrivial solutions \\
    \hline
  \end{tabular}
\end{table}

The paper is organized as follows. In section \ref{sec:basics}, we describe the basic geometry of the problem, the governing equations, and the form of solutions we are interested in finding. In section \ref{sec:viscous}, we analyze the Navier--Stokes equations in the case of constant nonzero viscosity. In section \ref{sec:inviscid}, we focus on the case of zero viscosity, governed by the Euler equations. Finally, in section \ref{sec:conclusions} we discuss the conclusions and implications of our findings for tornadogenesis. The appendix contains some auxiliary equations needed for our work that would unnecessarily clutter the presentation in the paper.

\section{Governing equations, basic geometry, and modified Serrin's variables}
\label{sec:basics}
In this section, we discuss the relevant equations, introduce a change of variables in the spirit of \cite{serrin}, and also introduce a special form of solutions we seek, which eventually allows us to reformulate the problem in terms of ordinary differential equations. Finally, we discuss the continuity equation and its implications in terms of boundary conditions.

\subsection{Governing equations}
The equations governing fluid flow are the Navier--Stokes equations,
\begin{equation}
\label{eq:ns}
  \rho\,\frac{D{\bf v}}{Dt}
  \equiv
  \rho\left(
        \frac{\partial{\bf v}}{\partial t}
        +
        ({\bf v}\cdot\nabla){\bf v}
      \right)
  =
  -\nabla P
  +
  (\lambda+\mu)\nabla(\nabla\cdot{\bf v})
  +
  \mu\,\Delta{\bf v},
\end{equation}
where $\bf v$, $P$, and $\rho$ are velocity, pressure, and density fields, respectively, and $\mu$ and $\lambda$ are dynamic viscosity coefficients.

The conservation of mass, or continuity, equation is
\begin{equation}
\label{eq:ce}
  \frac{D\rho}{Dt}+\rho\nabla\cdot{\bf v}
  \equiv
  \frac{\partial\rho}{\partial t}
  +
  {\bf v}\cdot\nabla\rho+\rho\,\nabla\cdot{\bf v}
  =
  0.
\end{equation}
We will consider the case of an incompressible and homogeneous flow, so that $D\rho/Dt=\nabla\cdot{\bf v}=0$ and $\nabla\rho=0$, respectively. We will seek steady-state solutions, i.e., those that satisfy ${\partial{\bf v}}/{\partial t}={\partial\rho}/{\partial t}=0$. Under these assumptions equation \eqref{eq:ce} is automatically satisfied. As a consequence, steady-state solutions for an incompressible, homogeneous flow satisfy the following simplified versions of \eqref{eq:ns} and \eqref{eq:ce},
\begin{equation}
\label{eq:nss}
  ({\bf v}\cdot\nabla){\bf v}
  =
  -\nabla p+\nu\,\Delta{\bf v}
\end{equation}
and
\begin{equation}
\label{eq:ces}
  \nabla\cdot{\bf v}
  =
  0,
\end{equation}
where $p = P/\rho$ is a (scaled) pressure field and $\nu=\mu/\rho$ is a (constant) kinematic viscosity. The relevant boundary conditions are
\begin{align*}
    {\bf v}                            = {\bf 0} & \qquad\text{when }\nu>0 \quad\text{(no source/sink and no slip)},\\
    \dfrac{\partial\bf v}{\partial n}  =0        & \qquad\text{when }\nu=0 \quad\text{(no source/sink, but slip allowed)}.
\end{align*}

\subsection{Spherical coordinate system and components of the velocity field}
Following Serrin \cite{serrin}, we use the (right) spherical coordinates $(R,\alpha,\theta)$, where $R$ is radial distance from the origin, $\alpha$ is the angle between the radius vector and the positive $z$-axis, and $\theta$ is the meridian angle about the $z$-axis. The positive $z$-axis then corresponds to $\alpha=0$ and the boundary (ground) plane to $\alpha=\pi/2$. We are interested in solutions in the upper half space, $z>0$, which corresponds to $R>0$ and $0\leq\alpha<\pi/2$ in our coordinate system.

We denote the components of the velocity vector ${\bf v}(R,\alpha,\theta)$ in the spherical coordinate system by $v_R$, $v_\alpha$, and $v_\theta$, and write
\begin{equation}
\label{eq:velRAT}
  {\bf v}(R,\alpha,\theta)
  =
  \left(
    v_R(R,\alpha,\theta),\,v_\alpha(R,\alpha,\theta),\,v_\theta(R,\alpha,\theta)
  \right).
\end{equation}
We will refer to the individual components as radial ($v_R$), meridional ($v_\alpha$), and azimuthal ($v_\theta$). The scaled pressure field will be denoted by $p(R,\alpha,\theta)$. The three components of the Navier--Stokes equations \eqref{eq:nss} in this coordinate system and in the spherical velocity components are given in the appendix in \eqref{eq:ns1}--\eqref{eq:ns3}, and the continuity equation \eqref{eq:ces} is given in \eqref{eq:ces1}.

We will follow Serrin's approach \cite{serrin} and consider velocities of the form
\begin{equation}
\label{eq:velFGO}
  v_R(R,\alpha,\theta)
  =
  \frac{G(x)}{r^b},
  \qquad
  v_\alpha(R,\alpha,\theta)
  =
  \frac{F(x)}{r^b},
  \qquad
  v_\theta(R,\alpha,\theta)
  =
  \frac{\Omega(x)}{r^b},
\end{equation}
where $r=R\sin\alpha$ is the horizontal distance to the $z$-axis, $x=\cos\alpha$, and $b>0$. We remark that the case studied by Serrin \cite{serrin} corresponds to $b=1$.

Since $\sin\alpha=\sqrt{1-x^2}$, we can use the change of variables
\begin{equation*}
  f(x)
  =
  F(x)(1-x^2)^{(1-b)/2},
  \qquad
  g(x)
  =
  G(x)(1-x^2)^{(1-b)/2},
  \qquad
  \omega(x)
  =
  \Omega(x)(1-x^2)^{(1-b)/2},
\end{equation*}
and rewrite \eqref{eq:velFGO} as
\begin{equation}
\label{eq:velfgo}
  v_R(R,\alpha,\theta)
  =
  \frac{g(x)}{R^b\sin\alpha},
  \qquad
  v_\alpha(R,\alpha,\theta)
  =
  \frac{f(x)}{R^b\sin\alpha},
  \qquad
  v_\theta(R,\alpha,\theta)
  =
  \frac{\omega(x)}{R^b\sin\alpha}.
\end{equation}
When $b=1$, the upper-case functions, $F$, $G$, and $\Omega$, agree with the lower-case functions, $f$, $g$, and $\omega$.

The continuity equation \eqref{eq:ces} written in terms of the newly introduced functions is given below in \eqref{eq:G-F}, while the Navier--Stokes equations \eqref{eq:nss} are discussed in section \ref{sec:nsFGO}.

We also remark that, as a consequence of \eqref{eq:G-F} below, the function
\begin{equation*}
  \Psi(R,\alpha)=R^{2-b}f(x)=R^{2-b}f(\cos\alpha)
\end{equation*}
satisfies
\begin{equation*}
  \nabla\Psi\cdot{\bf v}=0,
\end{equation*}
and therefore the surfaces $\Psi=\text{constant}$ contain the streamlines of the fluid motion. Notice that this is a direct generalization of Serrin's $\Psi=RF(x)$, since when $b=1$ we have $F(x)=f(x)$.

\subsection{The continuity equation}
We now consider the continuity equation \eqref{eq:ces} and its consequences for solutions to \eqref{eq:nss}. We first observe that direct substitution of \eqref{eq:velFGO} into the continuity equation \eqref{eq:ces} yields (see \eqref{eq:ces23} in the appendix)
\begin{equation}
\label{eq:G-F}
  \begin{split}
    (2-b)G(x)
    =
    \sqrt{1-x^2}\,&F'(x)
    -
    (1-b)\frac{x}{\sqrt{1-x^2}}F(x),\\
    (2-b)g(x)
    =\ &
    \sqrt{1-x^2}\,f'(x).
  \end{split}
\end{equation}
The prime symbol will denote differentiation with respect to $x$ throughout the paper. From \eqref{eq:G-F} we see that if $b\ne2$, then $G$ can be expressed in terms of $F$ and $g$ in terms of $f$.

We next derive an integral version of the continuity equation that will lead to naturally arising boundary conditions needed later in addition to the natural boundary condition that the ground contains no source or sink. Let $R_0>0$ and $E\subset\Real^3$ be the upper half ball bounded below by the horizontal disk $D=\{(R,\alpha,\theta):\ 0\le R<R_0,\ \alpha=\pi/2,\ 0\le\theta<2\pi\}$ and above by the hemisphere $S=\{(R,\alpha,\theta):\ R=R_0,\ 0\le\alpha<\pi/2,\ 0\le\theta<2\pi\}$, i.e.,
\begin{equation*}
  E
  =
  \{(R,\alpha,\theta):\ 0<R<R_0,\ 0\leq\alpha<\pi/2,\ 0\leq\theta<2\pi\}.
\end{equation*}
Applying \eqref{eq:ces} and the divergence theorem, we obtain
\begin{equation*}
  0
  =
  \iiint_E\nabla\cdot{\bf v}\,dV
  =
  \iint_{\partial E}{\bf v}\cdot{\bf n}\,dA
  =
  \iint_D v_\alpha\,dA
  +
  \iint_S v_R\,dA.
\end{equation*}
However, since there can be no source or sink at the ground ($\alpha=\pi/2$), we have that $v_\alpha(R,\pi/2,\theta)=F(0)/R^b=0$ for all $R>0$, or
\begin{equation}
  \label{eq:F0}
  F(0)
  =
  f(0)
  =
  0,
\end{equation}
and thus the integral over the disk $D$ vanishes. Substituting \eqref{eq:velFGO} into the integral over $S$, we obtain
\begin{equation*}
  \iint_S v_R\,dA
  =
  \int_0^{2\pi}\int_0^{\pi/2}\frac{G(x)}{r^b}R_0^2\sin{\alpha}\,d\alpha\,d\theta
  =
  2\pi R_0^{2-b}\int_0^{\pi/2}G(x)(\sin{\alpha})^{1-b}\,d\alpha
  =
  0,
\end{equation*}
or, in terms of $x$,
\begin{equation}
\label{eq:cont}
  \int_0^1\frac{G(x)}{(1-x^2)^{b/2}}\,dx
  =
  \int_0^1\frac{g(x)}{\sqrt{1-x^2}}\,dx
  =
  0.
\end{equation}
Substituting \eqref{eq:G-F} into \eqref{eq:cont}, integrating, and using the boundary value from \eqref{eq:F0}, we obtain
\begin{equation}
\label{eq:limitF}
  \lim_{x\to1}F(x)(1-x^2)^{(1-b)/2}
  =
  \lim_{x\to1}f(x)
  =
  0.
\end{equation}

\begin{remark}[Consequences of the continuity equation]
\label{rem:continuity}
We can summarize the consequences of the continuity equation \eqref{eq:ces} and the no source/sink boundary condition \eqref{eq:F0} as follows.
\begin{enumerate}
  \item If $b=2$, then, using \eqref{eq:G-F}--\eqref{eq:cont}, we have
    \begin{equation*}
      F=f\equiv0,
      \qquad
      \int_0^1\frac{G(x)}{1-x^2}\,dx=0.
    \end{equation*}
  \item If $b\ne2$, then, using \eqref{eq:F0}, \eqref{eq:limitF}, and \eqref{eq:G-F}, we have
    \begin{equation*}
      F(0)=\lim_{x\to1}F(x)(1-x^2)^{(1-b)/2}=0,
      \qquad
      (2-b)G(x)=\sqrt{1-x^2}\,F'(x)-(1-b)\frac{x}{\sqrt{1-x^2}}\,F(x),
    \end{equation*}
    or, in terms of the lower-case functions,
    \begin{equation*}
      f(0)=\lim_{x\to1}f(x)=0,
      \qquad
      (2-b)g(x)=\sqrt{1-x^2}\,f'(x).
    \end{equation*}
\end{enumerate}
\end{remark}
\begin{remark}
  Note that the special case $b=1$ in the previous remark gives rise to $F(0)=0$, $\lim_{x\to1}F(x)=0$, and $G(x)=\sqrt{1-x^2}\,F'(x)$, which is consistent with \cite{serrin}.
\end{remark}

\subsection{Simplification of the Navier--Stokes equations}
\label{sec:nsFGO}
In this section, we consider the Navier--Stokes equations \eqref{eq:nss} in terms of the velocity expressions \eqref{eq:velFGO} and \eqref{eq:velfgo}. We first observe that all of the velocity components have the form $v(R,\alpha,\theta)=K(\alpha)/R^b$, so their partial derivatives are of the form
\begin{equation*}
  \frac{\partial v}{\partial R}
  =
  -b\,\frac{K(\alpha)}{R^{b+1}},
  \qquad\qquad
  \frac{\partial v}{\partial\alpha}
  =
  \frac{\dot{K}(\alpha)}{R^b},
  \qquad\qquad
  \frac{\partial v}{\partial\theta}
  =
  0.
\end{equation*}
The dot symbol will denote differentiation with respect to $\alpha$ throughout the paper. Note that all terms in the left-hand sides of \eqref{eq:ns1}--\eqref{eq:ns3}, arising from the convective term in \eqref{eq:nss}, are of the form $C(\alpha)/R^{1+2b}$, while all terms in the right-hand sides, arising from the diffusive term in \eqref{eq:nss}, (i.e., all the terms multiplied by the viscosity coefficient, $\nu$) are of the form $D(\alpha)/R^{2+b}$. Therefore, the governing equations \eqref{eq:ns1}--\eqref{eq:ns3} have the general form
\begin{equation}
\label{eq:ns11}
  \frac{C_1(\alpha)}{R^{1+2b}}
  =
  -\frac{\partial p}{\partial R}
  +
  \nu\frac{D_1(\alpha)}{R^{2+b}},
\end{equation}
\begin{equation}
\label{eq:ns22}
  \frac{C_2(\alpha)}{R^{1+2b}}
  =
  -\frac{1}{R}\frac{\partial p}{\partial\alpha}
  +
  \nu\frac{D_2(\alpha)}{R^{2+b}},
\end{equation}
\begin{equation}
\label{eq:ns33}
  \frac{C_3(\alpha)}{R^{1+2b}}
  =
  -\frac{1}{R\sin\alpha}\frac{\partial p}{\partial\theta}
  +
  \nu\frac{D_3(\alpha)}{R^{2+b}},
 \end{equation}
where the expressions for $C_i(\alpha)$ and $D_i(\alpha)$ are given, in their various forms, in the appendix.

Like in \cite{serrin}, we argue that \eqref{eq:ns33} yields $\partial p/\partial\theta$ independent of $\theta$, so $p$ must be linear in $\theta$. Together with periodicity in $\theta$, this implies that $\partial p/\partial\theta=0$, so $p(R,\alpha,\theta)=p(R,\alpha)$. Consequently, \eqref{eq:ns33} reduces to
\begin{equation}
  \label{eq:ns333n}
  C_3(\alpha)
  =
  \nu R^{b-1}D_3(\alpha).
\end{equation}
The scaled pressure function has to satisfy \eqref{eq:ns11}, so by integrating it with respect to $R$ we obtain
\begin{equation*}
  p(R,\alpha)
  =
  \frac{C_1(\alpha)}{2bR^{2b}}
  -
  \nu\frac{D_1(\alpha)}{(1+b)R^{1+b}}
  +
  T(\alpha).
\end{equation*}
Substituting this expression into \eqref{eq:ns22}, we conclude that $\dot{T}(\alpha)=0$, and thus $T(\alpha)\equiv T$ is a constant and
\begin{equation}
\label{eq:pressure}
  p(R,\alpha)
  =
  \frac{C_1(\alpha)}{2bR^{2b}}
  -
  \nu\frac{D_1(\alpha)}{(1+b)R^{1+b}}
  +
  T.
\end{equation}
In addition, from \eqref{eq:ns22} and \eqref{eq:pressure} we obtain a {\it compatibility condition} for the existence of the pressure,
\begin{equation}
  \label{eq:ns12n}
  \dot{C}_1(\alpha)
  +
  2b\,C_2(\alpha)
  =
  \nu R^{b-1}\frac{2b}{1+b}\left(\dot{D}_1(\alpha)+(1+b)D_2(\alpha)\right).
\end{equation}
We now have the following equivalence lemma.
\begin{lemma}
  The system of Navier--Stokes equations \eqref{eq:ns11}--\eqref{eq:ns33} is equivalent to the system \eqref{eq:ns333n}--\eqref{eq:ns12n}.
\end{lemma}
\begin{proof}
  First, note that \eqref{eq:ns333n}--\eqref{eq:ns12n} follow directly from the Navier--Stokes equations \eqref{eq:ns11}--\eqref{eq:ns33}. Vice versa, if $C_i(\alpha)$ and $D_i(\alpha)$ are such that \eqref{eq:ns333n} and \eqref{eq:ns12n} are satisfied, then \eqref{eq:pressure} gives an expression for the scaled pressure so that \eqref{eq:ns11} is immediately satisfied, \eqref{eq:ns22} is satisfied due to the compatibility equation \eqref{eq:ns12n}, and \eqref{eq:ns33} follows immediately from \eqref{eq:ns333n}.
\end{proof}


\section{Analysis of the viscous case: $\nu>0$}
\label{sec:viscous}
In this section we discuss the existence of classical solutions to our problem with constant nonzero viscosity. We show that nontrivial solutions of the form \eqref{eq:velFGO} exist only for the case $b=1$ discussed by Serrin \cite{serrin}. We start by discussing the boundary conditions and then analyze the various cases that arise for various values of $b$.

\subsection{Boundary conditions}
In the case of nonzero viscosity, the no-slip requirement at the ground implies $v_R(R,\pi/2,\theta)=v_\theta(R,\pi/2,\theta)=0$ for all $R>0$ and $0\le\theta<2\pi$, so that, using \eqref{eq:velFGO}, $G(0)=\Omega(0)=0$. The no-sink/source requirement gives $v_\alpha(R,\pi/2,\theta)=0$, or \eqref{eq:F0}, $F(0)=0$. In addition, as a consequence of incompressibility, in particular \eqref{eq:G-F}, we have $F'(0)=0$. (Note that as discussed in Remark \ref{rem:continuity}, if $b=2$, then $F\equiv0$ and all boundary conditions concerning $F$ are automatically satisfied.)

Near the vortex axis, we have \eqref{eq:limitF} if $b\ne2$, while if $b=2$, then $F\equiv0$, and there are no physical restrictions on the behavior of $G$ and $\Omega$ as $x\to1$. However, we will assume, similarly as in \cite{serrin}, that near the vortex axis the azimuthal velocity, $v_\theta$, behaves like $C/r^b$, i.e., we will assume that
\begin{equation}
\label{eq:BC_omega}
  \lim_{x\to1}\Omega(x)
  =
  \lim_{x\to1}\omega(x)(1-x^2)^{-(1-b)/2}
  =
  C_\omega
  \ne
  0.
\end{equation}
Similarly as in \cite{serrin}, this boundary condition is based on the observation that ${\bf v}=\left(0,0,C_\omega/r^b\right)$ is a solution to our problem for any $b>0$ (see Sections \ref{sec:b=2} and \ref{sec:bne2} below), which can also be easily verified to be a solution in the full 3D space.

\begin{remark}[Boundary conditions in the viscous case]
\label{rem:BCviscous}
  In the case of constant nonzero viscosity, sought solutions $F$, $G$, and $\Omega$ are subject to the following requirements:
  \begin{itemize}
    \item If $b=2$, then
      \begin{equation*}
        F=f\equiv0,
        \qquad
        G(0)=\Omega(0)=g(0)=\omega(0)=0,
        \qquad
        \lim_{x\to1}\Omega(x)=C_\omega.
      \end{equation*}
    \item If $b\ne2$, then
      \begin{equation*}
        F(0)=F'(0)=G(0)=\Omega(0)=0,
        \qquad
        \lim_{x\to1}F(x)(1-x^2)^{(1-b)/2}=0,
        \qquad
        \lim_{x\to1}\Omega(x)=C_\omega.
      \end{equation*}
      or
      \begin{equation*}
        f(0)=f'(0)=g(0)=\omega(0)=0,
        \qquad
        \lim_{x\to1}f(x)=0,
        \qquad
        \lim_{x\to1}\omega(x)(1-x^2)^{-(1-b)/2}=C_\omega.
      \end{equation*}
  \end{itemize}
\end{remark}

\begin{remark}
  Note that the boundary conditions studied by Serrin are consistent with ours when $b=1$, since then $\lim_{x\to1}F(x)(1-x^2)^{(1-b)/2}=\lim_{x\to1}F(x)=0$.
\end{remark}

\subsection{Governing equations}
The governing equations are \eqref{eq:ns333n} and \eqref{eq:ns12n}, together with the continuity equations \eqref{eq:G-F}. We will need to distinguish between the special case $b=1$ studied by Serrin and the remaining cases when $b\ne1$.

\subsubsection{Case $b=1$}
In this case, \eqref{eq:ns333n} and \eqref{eq:ns12n} reduce to
\begin{equation*}
  C_3(\alpha)
  =
  \nu D_3(\alpha),
  \qquad
  \dot{C}_1(\alpha)+2\,C_2(\alpha)
  =
  \nu(\dot{D}_1(\alpha)
  +
  2D_2(\alpha)).
\end{equation*}
Using \eqref{eq:C3_FGO_n2}--\eqref{eq:D1D2_FGO_n2}, these equations can be rewritten as
\begin{equation}
\label{eq:serrin_5}
  \begin{split}
    \nu(1-x^2)F^{(4)}(x)
    -
    4\nu xF'''(x)
    +
    F(x)F'''(x)
    +
    3F'(x)F''(x)
    &=
    -\frac{2\Omega(x)\Omega'(x)}{1-x^2},\\
   \nu(1-x^2)\Omega''(x)
   +
   F(x)\Omega'(x)
   &=
   0.
  \end{split}
\end{equation}
We note that system \eqref{eq:serrin_5} is identical to system ($5$) in \cite{serrin} and is analyzed there. In what follows, we focus on the case with $b\ne1$.

\subsubsection{Case $b\ne1$}
In this case, the only way to satisfy \eqref{eq:ns333n} and \eqref{eq:ns12n} for all $R>0$ is to satisfy
\begin{equation}
\label{eq:bne1_NS}
  C_3(\alpha)
  =
  0,
  \qquad
  D_3(\alpha)
  =
  0,
  \qquad
  \dot{C}_1(\alpha)+2b\,C_2(\alpha)
  =
  0,
  \qquad
  \dot{D}_1(\alpha)+(1+b)D_2(\alpha)
  =
  0.
\end{equation}
The relevant expressions for the quantities in \eqref{eq:bne1_NS} are given in the appendix and we will recall them as needed. The last equation that needs to be satisfied is \eqref{eq:G-F}, the consequence of the continuity equation, restated here in both forms for completeness,
\begin{equation}
\label{eq:G-F2}
  (2-b)G(x)
  =
  \sqrt{1-x^2}\,F'(x)
  -
  (1-b)\frac{x}{\sqrt{1-x^2}}\,F(x)
  \qquad\text{ or }\qquad
  (2-b)g(x)
  =
  \sqrt{1-x^2}f'(x).
\end{equation}

\subsection{Case $b=2$ (no solutions)}
\label{sec:NSb=2}
We first address the special case with $b=2$. In this case, $F=f\equiv0$ and \eqref{eq:G-F2} provides no information. Using \eqref{eq:C3_FGO}--\eqref{eq:C1C2_FGO}, the first three equations in \eqref{eq:bne1_NS} reduce to
\begin{gather}
  G(x)\Omega(x)
  =
  0,
  \label{eq:b2_NSeqn1}\\
  (1-x^2)^2\Omega''(x)
  +
  2x(1-x^2)\Omega'(x)
  +
  3\Omega(x)
  =
  0,
  \label{eq:b2_NSeqn3}\\
  \frac{8x}{1-x^2}\,G^2(x)
  +
  2(G^2(x))'
  +
  (\Omega^2(x))'
  =
  0.
  \label{eq:b2_NSeqn2}
\end{gather}
Because of the boundary condition \eqref{eq:BC_omega}, $\Omega\ne0$ on some interval $(x_0,1)$ by continuity. Equation \eqref{eq:b2_NSeqn1} then implies that $G\equiv0$ in $(x_0,1)$, and \eqref{eq:b2_NSeqn2} reduces to $(\Omega^2)'=0$ in $(x_0,1)$. Thus $\Omega\equiv C_\omega$ in $(x_0,1)$. This, in turn, implies that $\Omega\equiv C_\omega$ and $G\equiv0$ in $(0,1)$. However, neither \eqref{eq:b2_NSeqn3}, nor the boundary condition $\Omega(0)=0$ are then satisfied, so no solution exists when $b=2$.

\subsection{Case $b\ne1,2$ (no solutions)}
\label{sec:NSbne12}
In this case, \eqref{eq:G-F2} can be substituted into \eqref{eq:bne1_NS} to yield the set of equations given in \eqref{eq:C3=0_G-F}--\eqref{eq:L1L2=0_G-F} in the appendix.

In order to conclude that no solutions exist, it suffices to analyze \eqref{eq:L3=0_G-F}, which reads
\begin{equation*}
  (1-x^2)^2\Omega''(x)
  -
  2(1-b)x(1-x^2)\Omega'(x)
  -
  (1-b^2)\Omega(x)
  =
  0.
\end{equation*}
When equipped with the initial conditions $\Omega(0)=0$ and $\Omega'(0)=C$, its solution is
\begin{equation}
\label{eq:hyperOmega}
  \Omega(x)
  =
  Cx(1-x^2)^{(b-1)/2}\,{}_2{\mathcal F}_1\left(\frac{1-b}{2},\frac{b}{2};\frac{3}{2};x^2\right),
\end{equation}
where ${}_2  \mathcal F_1$ is the Gaussian hypergeometric function given by (see \cite{NIST})
\begin{equation*}
 {}_2{\mathcal F}_1(\alpha,\beta;\gamma;z)
 =
 \sum_{n=0}^{\infty}\frac{(\alpha)_n(\beta)_n}{(\gamma)_n}\frac{z^n}{n!},
\end{equation*}
and $(x)_n$ with $n\in\N\cup\{0\}$ is the {\it Pochhammer symbol} given by
\begin{equation*}
  (x)_n
  =
  \begin{cases}
    1                  & \text{if }n=0,\\
    x(x+1)\dots(x+n-1) & \text{if }n>0.
  \end{cases}
\end{equation*}
Since the case $C=0$ in \eqref{eq:hyperOmega} would yield $\Omega\equiv0$, which does not satisfy the boundary condition \eqref{eq:BC_omega}, we only consider the case with $C\ne0$ and show that \eqref{eq:BC_omega} cannot be satisfied for any $b>0$.

We first have \cite[page 387, formula 15.4.20]{NIST}
\begin{equation*}
  {}_2{\mathcal F}_1\left(\frac{1-b}{2},\frac{b}{2};\frac{3}{2};1\right)
  =
  \frac{\Gamma\left(\frac{3}{2}\right)\Gamma(1)}{\Gamma\left(\frac{3-b}{2}\right)\Gamma\left(\frac{2+b}{2}\right)}
  =
  \frac{\sqrt{\pi}}{2\,\Gamma\left(\frac{3-b}{2}\right)\Gamma\left(\frac{2+b}{2}\right)}.
\end{equation*}
Since $1/\Gamma(z)$ is an entire function vanishing only for $z=0,-1,-2,\dots$, the value above is finite, and it is zero only for $b=3,5,7,\dots$ (or $b=-2,-4,-6,\dots$, but this case is excluded from our consideration). This observation, together with the behavior of $(1-x^2)^{(b-1)/2}$, allows us to conclude that
\begin{equation*}
  \lim_{x\to1-}\Omega(x)
  =
  \begin{cases}
    \infty & \text{if }0<b<1,\\
    0      & \text{if }b>1,
  \end{cases}
\end{equation*}
and the boundary condition \eqref{eq:BC_omega} cannot be satisfied for any choice of $C$. We can therefore conclude that no solutions with $b\ne1,2$ exist.

\section{Analysis of the inviscid case: $\nu=0$}
\label{sec:inviscid}
In this section we discuss the existence of classical solutions in case of zero viscosity. In tornadic thunderstorms one can expect a large Reynolds number on the order of $10^{10}$, and thus very small viscosity \cite{fiedlergarfield10}. When studying the case with $\nu=0$, we have to modify the boundary conditions at the ground and allow slip, and also tacitly assume that the solutions with $\nu=0$ are ``close'' to the physical solutions with large Reynolds numbers (see, e.g., \cite{dipernamajda87}).

We start this section by showing that the purely rotational {\it trivial} solution $F=G\equiv0$ and $\Omega\equiv C_\omega$ exists for all $b>0$, but that no nontrivial solutions of the form \eqref{eq:velFGO} exist if $b\ge2$. For $b=1$, we present and discuss analytic solutions and provide a numerical and graphical comparison with some of Serrin's solutions. We observe that our solutions with $b=1$ appear to be the limiting cases as viscosity tends to $0$, and, compared to our solutions, Serrin's solutions exhibit a boundary layer near the physical ground whose thickness tends to $0$. In particular, we observe that the size of the boundary layer tends to $0$ at the same rate as theoretically established in \cite{serrin}. For the cases $0<b<1$ and $1<b<2$, we present the governing equations that apparently admit nontrivial solutions. While we have not completed the existence and uniqueness analysis, we present our insights and numerical results in the case $0<b<1$, and we show that all potential solutions with $1<b<2$ could not satisfy Rayleigh's circulation criterion and would thus be unstable with respect to axisymmetric perturbations.

We again start by discussing the boundary conditions and then analyze the various cases that arise for various values of $b$.

\subsection{Boundary conditions}
Since slip is of no concern in the case of zero viscosity, we only need to address the no-source/sink requirements at the ground ($\alpha=\pi/2$) and at the center of the vortex ($\alpha=0$). From the analysis in the previous sections, it is clear that there will be no {\it a priori} restrictions on $G$ or $\Omega$. Regarding restrictions on $F$, we have
\begin{gather}
  F=f\equiv0
  \qquad\text{ if } b=2,\label{eq:BC_b=2}\\
  F(0)=f(0)=0
  \quad\text{ and }\quad
  \lim_{x\to1}F(x)(1-x^2)^{(1-b)/2}
  =
  \lim_{x\to1}f(x)
  =
  0
  \qquad\text{ if } b\ne2.\label{eq:BC_bnot2}
\end{gather}
We still assume that near the vortex axis the azimuthal velocity, $v_\theta$, behaves like $C/r^b$, i.e., we still assume that \eqref{eq:BC_omega} holds. We restate it here for completeness,
\begin{equation}
\label{eq:BC_omega2}
  \lim_{x\to1}\Omega(x)
  =
  \lim_{x\to1}\omega(x)(1-x^2)^{-(1-b)/2}
  =
  C_\omega
  \ne
  0.
\end{equation}

\subsection{Governing equations}
Since the viscosity coefficient, $\nu$, is zero, \eqref{eq:ns333n} and \eqref{eq:ns12n} reduce to
\begin{gather}
  C_3(\alpha)=0,\label{eq:C3=0}\\
  \dot{C}_1(\alpha)+2b\,C_2(\alpha)=0,\label{eq:C1C2=0}
\end{gather}
which can be rewritten using \eqref{eq:C3_FGO} and \eqref{eq:C1C2_FGO} in the appendix. The third equation is \eqref{eq:G-F} (restated later as \eqref{eq:G-F2}).

\begin{remark}
  Note that the governing equations \eqref{eq:nss} and \eqref{eq:ces} in the case of zero viscosity are invariant under the transformation $\bf{v}\mapsto-\bf{v}$, and so any obtained solution can be also ``reversed'' by changing its sign. In addition, we will see in some cases below that some of the equations are invariant under sign changes of some of the functions $F$, $G$, $\Omega$, etc., individually.
\end{remark}

\subsection{Case $b=2$ (no nontrivial solutions)}
\label{sec:b=2}
In this case, $F\equiv0$ by \eqref{eq:BC_b=2}, and \eqref{eq:G-F2} provides no information. Using \eqref{eq:C3_FGO} and \eqref{eq:C1C2_FGO}, equations \eqref{eq:C3=0} and \eqref{eq:C1C2=0} reduce to
\begin{gather}
  G(x)\Omega(x)
  =
  0,
  \label{eq:b2_eulereqn1}\\
  2
  \left[
    (G^2(x))'
    +
    \frac{4x}{1-x^2}\,G^2(x)
  \right]
  +
  (\Omega^2(x))'
  =
  0.
  \label{eq:b2_eulereqn2}
\end{gather}
Because of the boundary condition \eqref{eq:BC_omega2}, we have that $\Omega\ne0$ in some interval $(x_0,1)$ by continuity. Equation \eqref{eq:b2_eulereqn1} implies that $G\equiv0$ in $(x_0,1)$, and \eqref{eq:b2_eulereqn2} reduces to $(\Omega^2)'=0$ in $(x_0,1)$. Thus $\Omega\equiv C_\omega$ in $(x_0,1)$. This then implies that $\Omega\equiv C_\omega$ and $G\equiv0$ in $(0,1)$. Thus we obtain the ``trivial'' solution
\begin{equation}
\label{eq:trivial_sol}
  F=G\equiv0,
  \qquad
  \Omega\equiv C_\omega.
\end{equation}

\subsection{Case $b\ne2$ (existence of the trivial solution)}
\label{sec:bne2}
In this case, we can use relationship \eqref{eq:G-F2} between $G(x)$ and $F(x)$ and substitute it into \eqref{eq:C3=0} and \eqref{eq:C1C2=0}. The resulting equations are given in the appendix in \eqref{eq:C3=0_G-F} and \eqref{eq:C1C2=0_G-F}. Notice that we still have the trivial solution \eqref{eq:trivial_sol},
since if $\Omega\equiv C_\omega$, then \eqref{eq:C3=0_G-F} implies $F(x)=c\sqrt{1-x^2}$, and the initial condition $F(0)=0$ gives $c=0$ and $F\equiv0$; equation \eqref{eq:G-F2} then gives $G\equiv0$.

There remains to be seen if there exist other, nontrivial solutions. We first address the simple case $b=1$ and then turn to the more complicated case $b\ne1,2$.

\subsection{Case $b=1$ (existence of nontrivial solutions)}
\label{sec:b=1}
If $b=1$, then \eqref{eq:C3=0_G-F} and \eqref{eq:C1C2=0_G-F} reduce to
\begin{gather}
  F(x)\Omega'(x)=0,\label{eq:b1_eulereqn1}\\
  (\Omega^2(x))'+\frac{1}{2}(1-x^2)\left(F^2(x)\right)'''=0.\label{eq:b1_eulereqn2}
\end{gather}
Recall that in this case $F$ vanishes at both $x=0$ and $x=1$ by \eqref{eq:BC_bnot2}. We can consider two cases. Either $F\equiv0$, in which case \eqref{eq:b1_eulereqn2} implies $\Omega\equiv C_\omega$, \eqref{eq:b1_eulereqn1} is trivially satisfied, and from \eqref{eq:G-F2} we have $G\equiv0$. This case corresponds to the trivial solution \eqref{eq:trivial_sol}.

In the second case, if $F(x_0)\ne0$ for some $x_0\in(0,1)$, then consider the largest interval $(x_1,x_2)\subset(0,1)$ containing $x_0$ such that $F(x)\ne0$ in $(x_1,x_2)$ and $F(x_1)=F(x_2)=0$. In $(x_1,x_2)$, $\Omega$ has to be constant in view of \eqref{eq:b1_eulereqn1} and $F^2(x)=c(x-x_1)(x_2-x)$ with $c>0$ in view of \eqref{eq:b1_eulereqn2}. However, in this case all (one-sided) derivatives of $F$ become infinite at $x_1$ and $x_2$, and therefore the only possibility is that $x_1=0$ and $x_2=1$, in which case we have $\Omega\equiv C_\omega$, $F(x)=C_1\sqrt{x(1-x)}$ with $C_1\ne0$, and, from \eqref{eq:G-F2}, $G(x)=C_1\dfrac{(1-2x)\sqrt{1+x}}{2\sqrt{x}}$.

In summary, when $b=1$, we have a set of solutions of the form
\begin{equation}
\label{eq:sol_b=1}
  \Omega\equiv C_\omega,
  \qquad
  F(x)=C_1\sqrt{x(1-x)},
  \qquad
  G(x)=C_1\frac{(1-2x)\sqrt{1+x}}{2\sqrt{x}}
  \quad\text{ for }
  C_1\in\Real,
\end{equation}
which also includes the trivial solution \eqref{eq:trivial_sol} when $C_1=0$.

We see that the solutions with $C_1\ne0$ will have infinite flow speeds both near the ground and near the vortex axis. The velocity becomes infinite near the ground in the radial direction (inflow for updraft solutions with $C_1<0$ and outflow for downdraft solutions with $C_1>0$) due to $G$ having an asymptote at $x=0$. Near the vortex axis ($x=1$) both $\Omega$ and $G$ have a finite limit and therefore the flow speed becomes infinite due to the $r$ term in the denominators of the velocity components \eqref{eq:velFGO}. Both of these phenomena are observed in Fig.~\ref{fig:b=1} in the middle plot.

\begin{remark}
  Note that we are only looking for classical solutions for $x\in(0,1)$ that lead to the solution \eqref{eq:sol_b=1}, resulting in only updraft or only downdraft flows. If we allowed more general solutions, we could generate flows with an arbitrary number of cells, $n$, with alternating updraft and downdraft flows by considering a partition of the interval $(0,1)$, say, $0=a_0<a_1<\dots<a_n=0$, and on each $(a_i,a_{i+1})$ have $F^2(x)=c_i(x-a_i)(a_{i+1}-x)$ with $c_i>0$. By alternating the signs of $F$ from interval to interval, we could obtain $\lim_{x\to a_i}F'(x)=\lim_{x\to a_i}G(x)=+\infty$ or $-\infty$ for each $0<i<n$, and thus obtain a collection of conical flows with infinite inflows or outflows along every cone.
\end{remark}

To assess how reasonable solution \eqref{eq:sol_b=1} is, we have implemented the iterative procedure described by Serrin \cite{serrin} to compute solutions to \eqref{eq:serrin_5} with small nonzero viscosity. In \cite{serrin}, solutions depend on two parameters, $k$ and $P$; viscosity is related to $k$ via $\nu=1/(2k)$, so small values of viscosity correspond to large values of $k$. In Fig.~\ref{fig:serrin_solution} we present two solutions, one for $k=100$ and $P=0$ (left) and one for $k=1000$ and $P=0$ (right), and we compare them to \eqref{eq:sol_b=1} with $C_1=C_\omega=1$. We observe very good agreement of the two solutions in the interval $(0,1)$ except near $x=0$, where the nonzero-viscosity solution exhibits a thin boundary layer due to the no-slip boundary conditions, while $F'$ and $G$ from \eqref{eq:sol_b=1} both tend to infinity as $x\to0$. Increasing $k$ (i.e., decreasing the viscosity, $\nu$) results in shrinking of the size of the boundary layer. More specifically, to numerically estimate the size of this boundary layer, we focus on the $x$-value at which Serrin's $\Omega(x)$ (blue curve in Fig.~\ref{fig:serrin_solution}) starts to deviate from the inviscid solution $\Omega(x)\equiv1$ (dashed). For several decreasing values of $\nu$ we estimate the layer size and plot the results on a log-log scale in Fig.~\ref{fig:bdry_layer}. We observe a linear relationship between the logarithm of the layer size and the logarithm of the viscosity with a slope estimated by linear regression to be approximately $0.669$. In \cite{serrin}, Serrin defines the boundary layer independently of the solutions of Euler equations, and analytically estimates its size to be on the order of $\nu^{2/3}$, which very well agrees with our result. We, therefore, conclude that outside this boundary layer the solutions of Navier--Stokes equations are in good agreement with the solutions of Euler equations.

This provides numerical evidence that downdraft solutions ($C_1>0$) in \eqref{eq:sol_b=1} are limits of downdraft solutions of \eqref{eq:serrin_5} as $\nu\to0$. On the other hand, it follows from Serrin's results that updraft solutions ($C_1<0$) in \eqref{eq:sol_b=1} cannot be limits as viscosity tends to zero of any of the solutions presented in \cite{serrin}. This leaves open the question whether solutions other than those described by Serrin exist that tend to solution \eqref{eq:sol_b=1} with $C_1<0$ as viscosity tends to zero.
\begin{figure}
  \includegraphics[width=0.45\textwidth]{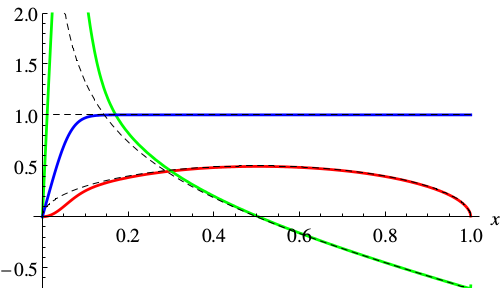}
  \hfill
  \includegraphics[width=0.45\textwidth]{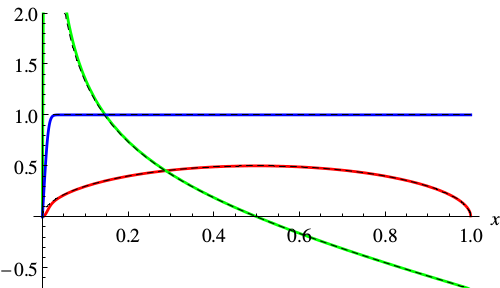}
  \caption{Comparison of the inviscid flow solution \eqref{eq:sol_b=1} with $C_1=C_\omega=1$ with the solutions with small nonzero viscosity corresponding to $k=100$ and $P=0$ (left) and $k=1000$ and $P=0$ (right) in \cite{serrin}. The plot of $F$ is shown in red, plot of $G$ in green, and plot of $\Omega$ in blue, with the plots for solution \eqref{eq:sol_b=1} dashed.}
  \label{fig:serrin_solution}
\end{figure}

\begin{figure}
  \includegraphics[width=0.55\textwidth]{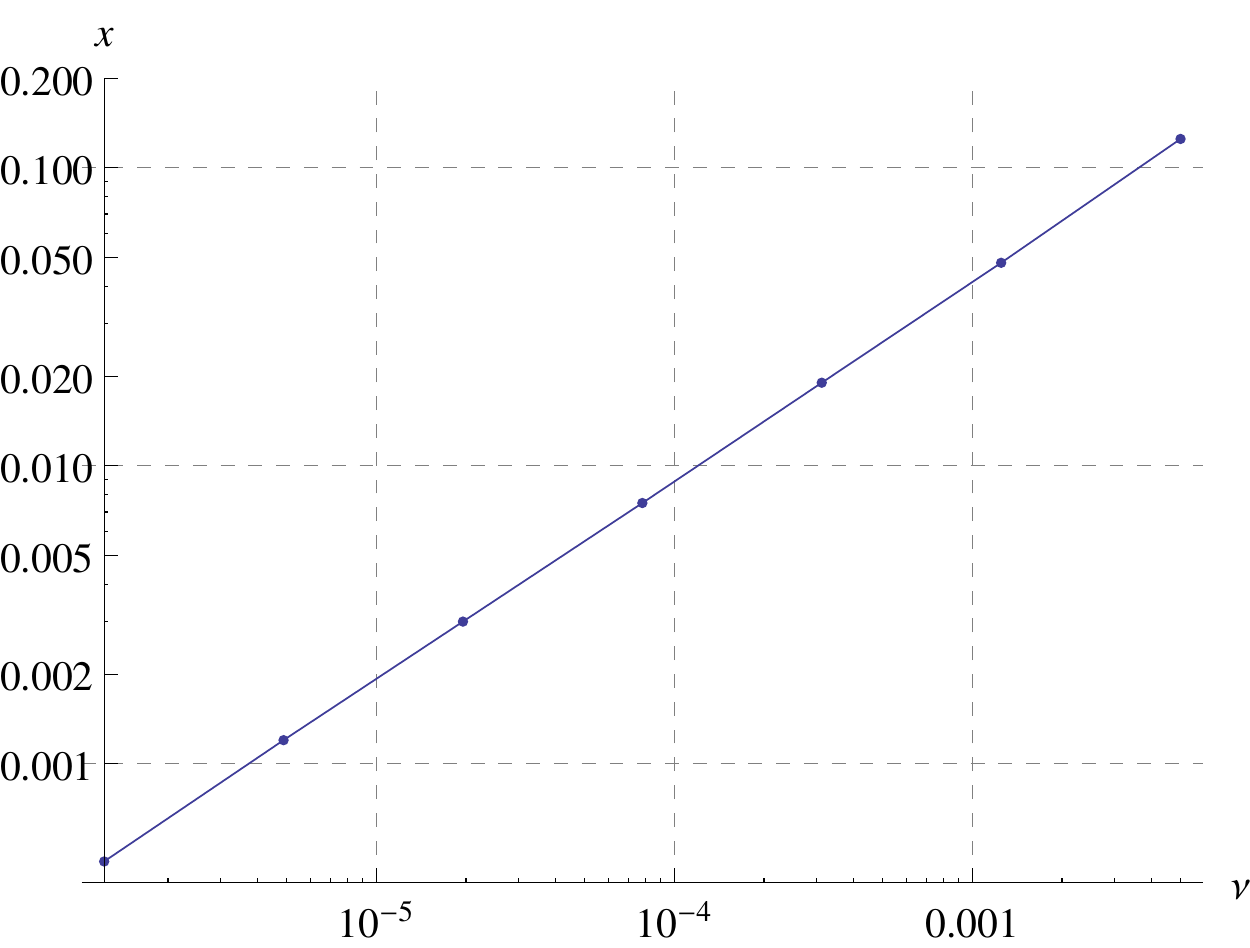}
  \caption{Boundary layer analysis for the case $b=1$. A linear relationship between the logarithm of the kinematic viscosity (horizontal axis) and the logarithm of the estimated size of the boundary layer (vertical axis) is observed with a slope of $0.669$ obtained by linear regression.}
  \label{fig:bdry_layer}
\end{figure}

To visualize solution \eqref{eq:sol_b=1} in other ways, in Fig.~\ref{fig:b=1} we show a streamlines plot that represents particle trajectories without the azimuthal component, a contour plot of the speed, $\|{\bf v}\|=\sqrt{v_R^2+v_\alpha^2+v_\theta^2}$, and a contour plot of the pressure obtained from \eqref{eq:pressure} with $T=0$. The shown ranges are $0<r<1$, $0<z<1$ with $r=R\sin\alpha$ and $z=R\cos\alpha$. We choose $C_1=4\sqrt{2}$ and $C_\omega=1$ since the corresponding solution is also displayed later in red in Fig.~\ref{fig:FOmegaG_c=0.25} (as a limit of numerically computed solutions corresponding to $b\nearrow1$). All of the contour plots in this paper have been generated with fifty, automatically chosen and uniformly spaced contour levels. Due to the singularities in the speed and the pressure near the axis of the vortex or near the ground, the holes that appear in the isospeed and isobar plots correspond to values that are out of the automatically chosen range.
\begin{figure}
  \centering
  \includegraphics[width=0.32\textwidth]{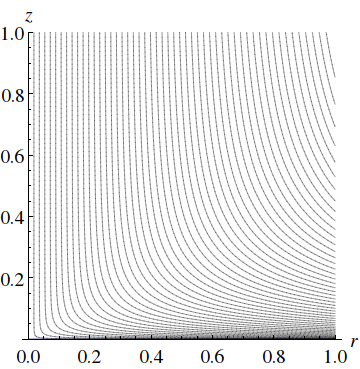}
  \hfill
  \includegraphics[width=0.32\textwidth]{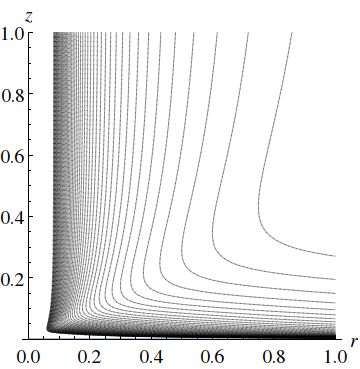}
  \hfill
  \includegraphics[width=0.32\textwidth]{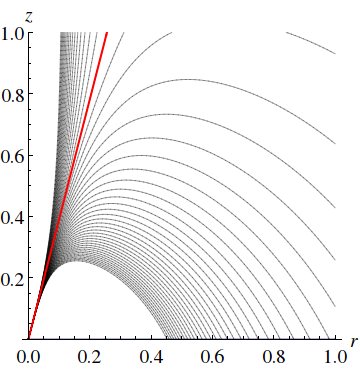}
  \caption{Plots of the streamlines (left), the corresponding isocurves for speed (middle), and isocurves for pressure (right) for the solution in \eqref{eq:sol_b=1} with $C_1=4\sqrt{2}$ and $C_\omega=1$. The horizontal axes correspond to $r=R\sin\alpha$ and the vertical axes to $z=R\cos\alpha$. The values for speed range from $4$ to $50$ with a step of $1$, increasing towards the vortex axis. The values for pressure range from $-36.8$ to $75.0$ with a step of $2.3$ with lowest values near the vortex axis. The straight red line corresponds to the level set $p(R,\alpha)=0$, which is the line $\cos\alpha=1-(C_\omega/C_1)^2$.}
  \label{fig:b=1}
\end{figure}

We remark that the solution in \eqref{eq:sol_b=1} is self-similar, and the self-similarity is clearly observed when one zooms out of the plots in Fig.~\ref{fig:b=1}. The zoomed out figures look identical to Fig.~\ref{fig:b=1} with the contours only corresponding to different values for each level set. This is clear from the definition of the velocities in \eqref{eq:velFGO}, since the functions $F$, $G$, and $\Omega$ only depend on $x=\cos\alpha$. As an illustration of Cai's power law method, we computed the exponent for the velocity-radius power law from the data in the middle plot in Fig.~\ref{fig:b=1} by computing the slope of the logarithm of the speed against the logarithm of the scale for several different pairs of points and obtained $-1$ for the slope. This computation was done using the values at the height of $1.0$ unit in the middle plot, where the speed values can be easily read off. While this height was chosen for convenience, the results would be the same at any height due to the assumption on the structure of the solution \eqref{eq:velFGO}.

The pressure plot in Fig.~\ref{fig:b=1} shows low values near the vortex axis and finite values along the ground, which increase as $r\to0$. In fact, from the solution \eqref{eq:sol_b=1} and the expression for pressure \eqref{eq:pressure}, one quickly obtains (taking $T=0$) that
\begin{equation*}
  p(R,\alpha)
  =
  -\frac{C_\omega^2-C_1^2(1-x)}{2r^2}
  =
  -\frac{C_\omega^2-C_1^2(1-\cos\alpha)}{2R^2\sin^2\alpha},
\end{equation*}
so we see that as one approaches the vortex axis, i.e., as $x\to1$, the pressure behaves like $-1/r^2$. (Note that the physical pressure has the form $p(R,\alpha)+T$ and thus has a singularity near the vortex axis no matter what the value of $T$ is. As in Serrin's approach, this is a consequence of the assumption on the velocity \eqref{eq:velFGO}.) It is also immediate to observe that the pressure is zero along the line $x=1-(C_\omega/C_1)^2$; this line is indicated by the bolder red line in Fig.~\ref{fig:b=1}. Notice that near the corner, where the vortex axis meets the ground, all of the other level curves are tangent to this line,
and our model in this case would formally indicate a large pressure gradient (singular at the origin as well as along the vortex axis)
as the pressure undergoes a sudden change from positive to negative values when crossing the red line and approaching the vortex axis. Clearly, such a behavior will be observed if $C_1^2>C_\omega^2$, since then the level line $p(R,\alpha)=0$ have positive slope. In Fig.~\ref{fig:b=1_others}, we show the cases that correspond to the line $p(R,\alpha)=0$ having angles $\pi/4$, $0$, and $-\pi/4$ with the horizontal, respectively. The corresponding values of $C_1^2$ are then $2+\sqrt{2}$, $1$, and $2-\sqrt{2}$, respectively (with $C_\omega=1$). We note that the apparent singularity of the pressure gradient near the vortex axis might be due to the original assumption on the velocity field \eqref{eq:velFGO} and therefore not be physically reasonable.
\begin{figure}
  \centering
  \includegraphics[width=0.32\textwidth]{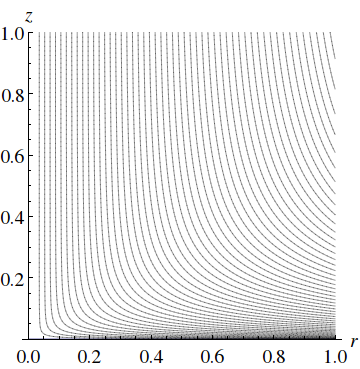}
  \hfill
  \includegraphics[width=0.32\textwidth]{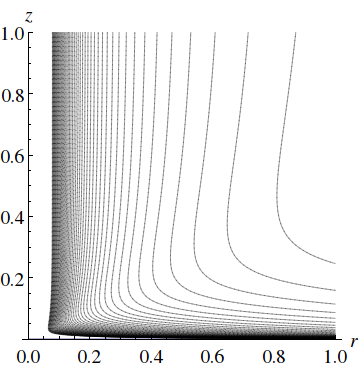}
  \hfill
  \includegraphics[width=0.32\textwidth]{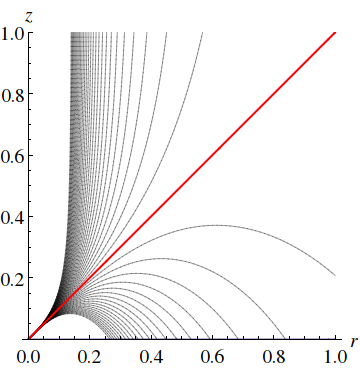}
  \vskip\baselineskip
  \includegraphics[width=0.32\textwidth]{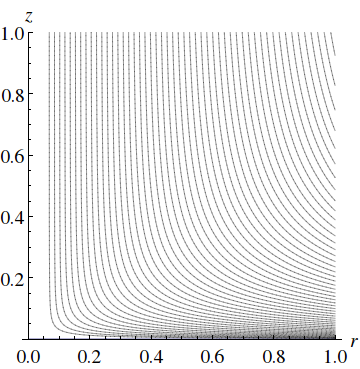}
  \hfill
  \includegraphics[width=0.32\textwidth]{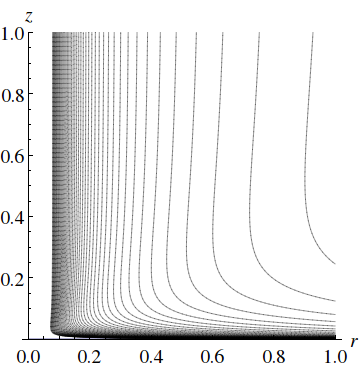}
  \hfill
  \includegraphics[width=0.32\textwidth]{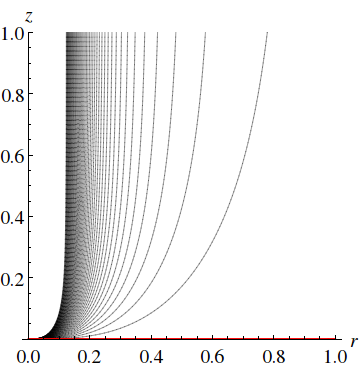}
  \vskip\baselineskip
  \includegraphics[width=0.32\textwidth]{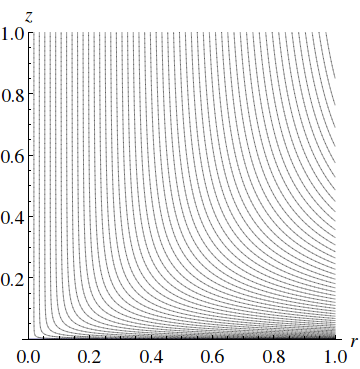}
  \hfill
  \includegraphics[width=0.32\textwidth]{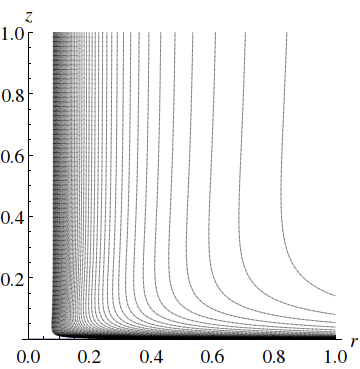}
  \hfill
  \includegraphics[width=0.32\textwidth]{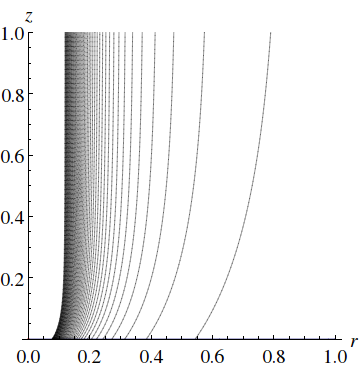}
  \caption{Plots of the streamlines (left), the corresponding isocurves for speed (middle), and isocurves for pressure (right) for the solution in \eqref{eq:sol_b=1} with $C_\omega=1$ and $C_1=\sqrt{2+\sqrt{2}}$ (top row), $C_1=1$ (middle row), and $C_1=\sqrt{2-\sqrt{2}}$ (bottom row). The values for speed range from $1.68$ to $21.00$ with a step of $0.42$ (top), from $1.24$ to $15.50$ with a step of $0.31$ (middle), and from $1.30$ to $13.78$ with a step of $0.26$ (bottom), increasing towards the vortex axis. The values for pressure range from $-24.94$ to $17.20$ with a step of $0.86$ (top), from $-32.50$ to $0$ (red line) with a step of $0.65$ (middle), and from $-0.7$ to $-35.0$ with a step of $0.7$ (bottom), with lowest values near the vortex axis.}
  \label{fig:b=1_others}
\end{figure}

The pressure plots in Fig.~\ref{fig:b=1} and Fig.~\ref{fig:b=1_others} also provide an interesting characterization of possible shapes of a visible tornado funnel. Since the funnel outline should approximately follow the isobars, we see three distinct possible shapes: one that is conical near the ground (Fig.~\ref{fig:b=1} and Fig.~\ref{fig:b=1_others} (top row)); one that can be viewed as a degenerate, fully open cone, yet still with a single point touching the ground (Fig.~\ref{fig:b=1_others} (middle row)); and one with the funnel having a nonzero width at the ground (Fig.~\ref{fig:b=1_others} (bottom row)). Note that changing $C_1$ while keeping $C_\omega$ fixed changes the relative magnitudes of the azimuthal component of the velocity with respect to the non-azimuthal ones. We could view a large value of $C_\omega/C_1$ as corresponding to large amount of swirl, and a small value corresponding to a small amount of swirl \cite{ward72,davies-jones73,lewellensxia00}. In this sense, we can say that in Fig.~\ref{fig:b=1_others} swirl increases from top to bottom, and wider funnels correspond to more swirl. We illustrate this behavior in Fig.~\ref{fig:b=1_swirl}, in which a few streamlines are shown for three cases, $C_1=-10$ (left), $C_1=-1$, and $C_1=-1/10$ in \eqref{eq:sol_b=1}. It is believed that increasing the swirl ratio in a single-cell vortex can lead to a vortex breakdown into multiple vortices as shown in Fig.~\ref{fig:breakdown} \cite{rotunno13}. Our model only captures an updraft or a downdraft flow, so even though the swirl increases from top to bottom in Fig.~\ref{fig:b=1_others}, our model cannot capture the whole dynamics of a vortex breakdown. See the conclusions section for more discussion of a vortex breakdown.
\begin{figure}
  \centering
  \includegraphics[width=0.32\textwidth]{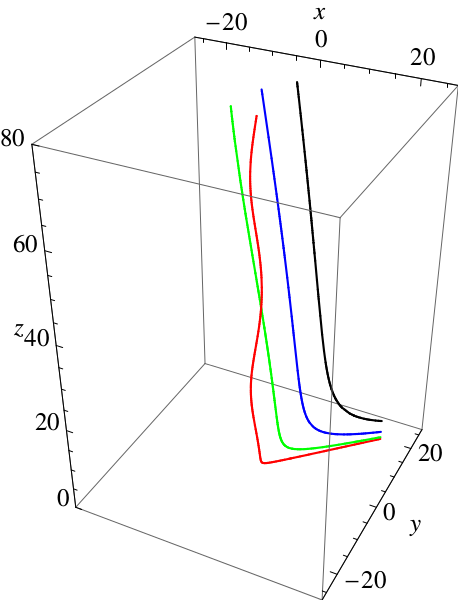}
  \hfill
  \includegraphics[width=0.32\textwidth]{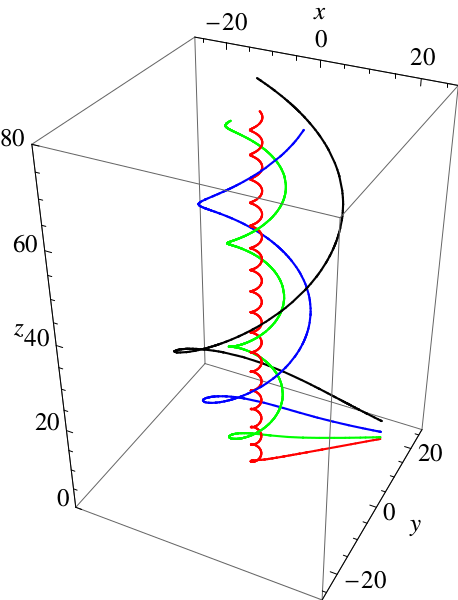}
  \hfill
  \includegraphics[width=0.32\textwidth]{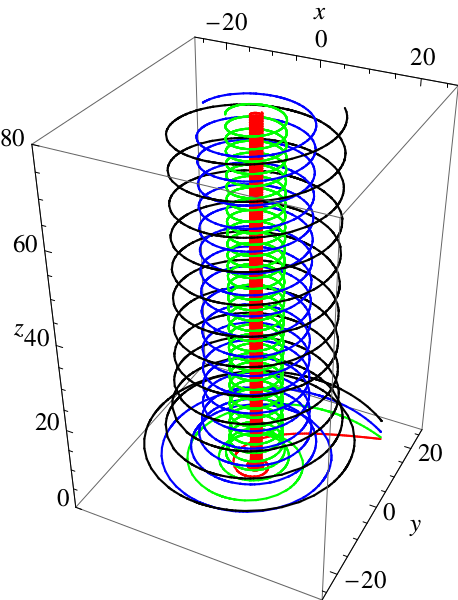}
  \caption{Illustration of the effect of the ratio $C_\omega/C_1$ on the swirl of solutions \eqref{eq:sol_b=1}. In all cases $C_\omega=1$, but $C_1=-10$ in the left plot, $C_1=-1$ in the middle plot, and $C_1=-1/10$ in the right plot. Four streamlines are plotted in each case, with initial points $(20,20,0.02)$, $(20,20,0.5)$, $(20,20,2)$, and $(20,20,5)$.}
  \label{fig:b=1_swirl}
\end{figure}

\subsection{Case $b\ne1,2$ (general observations)}
In this case, the governing equations \eqref{eq:C3=0} and \eqref{eq:C1C2=0}, using the lower-case functions and expressing $g$ using \eqref{eq:G-F}, reduce to
\begin{equation}
\label{eq:C3=0_bne1}
  f\omega'
  =
  \frac{1-b}{2-b}\,f'\omega,
\end{equation}
\begin{equation}
\label{eq:C1C2=0_bne1}
  (1-x^2)
  \left[
    \frac{2+b}{2-b}\,f'f''
    +
    ff'''
  \right]
  +
  4(1-b)ff'
  +
  2(1-b)(2-b)
  \frac{x}{1-x^2}
  \left(
    f^2+\omega^2
  \right)
  +
  2(2-b)\omega\omega'
  =
  0.
\end{equation}

\begin{remark}
The last two terms containing $\omega$ in \eqref{eq:C1C2=0_bne1} can be written in terms of $\Omega$ in the form
\begin{equation*}
  2(2-b)
  \left[
    \omega\omega'
    +
    (1-b)\frac{x}{1-x^2}\,\omega^2
  \right]
  =
  2(2-b)(1-x^2)^{1-b}\,\Omega\Omega',
\end{equation*}
and hence we see from \eqref{eq:C1C2=0_bne1} that if $F=f\equiv0$, then we recover the trivial solution \eqref{eq:trivial_sol}, $F=G\equiv0$ and $\Omega\equiv C_\omega$. Similarly, if $\Omega\equiv C_\omega$, then it follows from \eqref{eq:C3=0_bne1} and the boundary condition $F(0)=f(0)=0$ that $F=f\equiv0$. We thus seek solutions with $f\not\equiv0$ (or $F\not\equiv0$) and $\Omega\not\equiv C_\omega$.
\end{remark}

It will be more convenient to work in terms of $f$ and $\Omega$, so we rewrite equations \eqref{eq:C3=0_bne1} and \eqref{eq:C1C2=0_bne1} as
\begin{equation}
\label{eq:C3=0_fOmega}
  f\Omega'
  =
  \frac{1-b}{2-b}
  \left[
    f'
    +
    (2-b)\frac{x}{1-x^2}\,f
  \right]
  \Omega,
\end{equation}
\begin{equation}
\label{eq:C1C2=0_fOmega}
  (1-x^2)
  \left[
    \frac{2+b}{2-b}\,f'f''
    +
    ff'''
  \right]
  +
  4(1-b)ff'
  +
  2(1-b)(2-b)
  \frac{x}{1-x^2}
  f^2
  +
  2(2-b)(1-x^2)^{1-b}\,\Omega\Omega'
  =
  0.
\end{equation}
\begin{remark}
  Notice that both \eqref{eq:C3=0_fOmega} and \eqref{eq:C1C2=0_fOmega} are invariant under sign changes $f\mapsto-f$ and $\Omega\mapsto-\Omega$; as one consequence, we can assume that $C_\omega>0$.
\end{remark}

On any interval where $f\ne0$, we can solve \eqref{eq:C3=0_bne1} for $\omega$ and obtain
\begin{equation}
\label{eq:sol_omega_f}
  \omega(x)
  =
  c|f(x)|^{(1-b)/(2-b)}
  \quad\text{ for }c>0.
\end{equation}
Vice versa, on any interval where $\omega\ne0$ (and in particular on some interval $(x_1,1)$ due to the boundary condition \eqref{eq:BC_omega2}) we can solve for $f$ and obtain
\begin{equation}
\label{eq:sol_f_omega}
  f(x)
  =
  c|\omega(x)|^{(2-b)/(1-b)}
  =
  c(1-x^2)^{(2-b)/2}|\Omega(x)|^{(2-b)/(1-b)}
  \quad\text{ for }c>0.
\end{equation}
We thus have that
\begin{equation}
\label{eq:f_asymptotics}
  f(x)
  =
  \mathcal{O}\left((1-x^2)^{(2-b)/2}\right)
  \qquad\text{and}\qquad
  F(x)
  =
  \mathcal{O}\left(\sqrt{1-x^2}\right)
  \qquad\text{ as }x\to1.
\end{equation}

\subsection{Case $2<b<\infty$ (no nontrivial solutions)}
\label{sec:b>2}
We now show that no nontrivial solutions of \eqref{eq:C3=0_fOmega} and \eqref{eq:C1C2=0_fOmega} in the classical sense can exist for $b>2$. First, from the definition of $\omega(x)=\Omega(x)(1-x^2)^{(1-b)/2}$ and the boundary condition \eqref{eq:BC_omega2} we see that $\omega(x)\to+\infty$ as $x\to1$. Since in this case $\frac{2-b}{1-b}>0$, then, in view of \eqref{eq:sol_f_omega}, the only way to not violate the boundary condition $f(x)\to0$ as $x\to1$ is if $f$ is identically zero on some interval $[x_1,1)$. Since we are looking for solutions with $f\not\equiv0$, let us assume, without loss of generality, that $f>0$ on some interval $(x_0,x_1)$. We now compare the limiting behavior of $\omega$ on either side of $x_1$. On $(x_1,1)$, where $f\equiv0$, we have from \eqref{eq:C1C2=0_fOmega} that $\Omega\equiv C_\omega$, so $\omega$ has a nonzero limit as $x\to x_1$ from the right. On the other hand, on $(x_0,x_1)$, where $f>0$, we have \eqref{eq:sol_omega_f}, and hence $\lim_{x\to x_1-}\omega(x)=0$. Therefore $\omega$ cannot be continuous and there are no nontrivial solutions in the case $2<b<\infty$.

\subsection{Case $1<b<2$ (behavior of potential solutions)}
In this case we have $\frac{2-b}{1-b}<0$. Since again $\omega(x)\to+\infty$ as $x\to1$, we can use \eqref{eq:sol_f_omega} to express $f$ in terms of $\omega$ and observe that this time the boundary condition $f(x)\to0$ as $x\to1$ is satisfied independently of the value of $c$ in \eqref{eq:sol_f_omega}. Note that in this case $\omega$ has to be positive in $(0,1)$, since if $\omega(x_0)=0$ for some $x_0\in(0,1)$, then, in view of \eqref{eq:sol_f_omega}, $f$ would have an asymptote at $x_0$ and thus be discontinuous there. It follows from \eqref{eq:sol_f_omega} that $f$ cannot change sign in $(0,1)$ either, and \eqref{eq:sol_omega_f} implies that $\omega(x)\to+\infty$ as $x\to0$. Consequently, we also have $\Omega(x)\to+\infty$ as $x\to0$ and the azimuthal velocity becomes infinite near the ground.

We have been unable to find analytic expressions for such solutions and also encountered difficulties when approximating them numerically. However, in the next section, we show that such solutions would be unstable with respect to axisymmetric perturbations.

\subsection{Instability of potential solutions for $1<b<2$}
\label{sec:b>1}
In this section we will assume $0<b<2$ and address the centrifugal stability of solutions to \eqref{eq:C3=0_fOmega} and \eqref{eq:C1C2=0_fOmega} with respect to axisymmetric perturbations. We will use Rayleigh's circulation criterion \cite{drazinreid}, which can be stated as the requirement that the Rayleigh discriminant $\Phi$ is nonnegative, where
\begin{equation*}
  \Phi(r)
  =
  \frac{1}{r^3}\frac{\partial}{\partial r}(rv_\theta)^2.
\end{equation*}
Substituting in the expression $v_\theta=\Omega(x)/r^b$ and using the relationship between $x$ and the cylindrical coordinates $r$ and $z$, $x=\cos\alpha=z/\sqrt{r^2+z^2}$, we obtain
\begin{equation}
\label{eq:Phi}
  \Phi(r)
  =
  \frac{2}{r^{2(1+b)}}\,\Omega(x)\left[(1-b)\Omega(x)-x(1-x^2)\Omega'(x)\right].
\end{equation}
The particular case with $b=1$ gives $\Phi\equiv0$ since from \eqref{eq:sol_b=1} we have $\Omega\equiv C_\omega$. Also, the trivial solution with $F=G\equiv0$ and $\Omega\equiv C_\omega$ is clearly stable only for $0<b\le1$. We will show that $b=1$ is the largest value of $b$ that permits stable nontrivial solutions.

For $1<b<2$, the stability requirement $\Phi(r)\ge0$ can be replaced by an equivalent statement $f^2(x)\Phi(r)\ge0$ since $f\ne0$ in $(0,1)$. We can then use \eqref{eq:C3=0_fOmega} to get
\begin{equation*}
  f^2(x)\Phi(r)
  =
  \frac{2(1-x^2)\Omega^2(x)}{r^{2(1+b)}}\frac{1-b}{2-b}f(x)
  \left[
    (2-b)f(x)
    -
    xf'(x)
  \right],
\end{equation*}
which is invariant under the sign change of $f$, so we can assume, without loss of generality, that $f>0$ in $(0,1)$. The stability requirement $f^2(x)\Phi(r)\ge0$ now implies $(2-b)f(x)-xf'(x)\le0$, and, in particular, $f$ is nondecreasing in $(0,1)$. However, this contradicts the assumptions $f>0$ in $(0,1)$ and $\lim_{x\to1}f(x)=0$, so no solutions corresponding to $1<b<2$ can be stable with respect to axisymmetric perturbations.

\subsection{Case $0<b<1$ (numerical solutions)}
\label{sec:b<1}
Lack of analytic solutions for $b=1$ and constant nonzero viscosity led to numerical approaches presented, e.g., in \cite{serrin,hamada07,madani03}. We have not been able to find analytic expressions for any nontrivial solutions in the case $0<b<1$ either, but we used a numerical approach to generate their approximations for various values of $b$ between $0$ and $1$. Once our solutions are computed, it can then be numerically or graphically verified that they satisfy the stability criterion $\Phi\ge0$ with $\Phi$ given in \eqref{eq:Phi}. All of our numerical solutions for $0<b<1$ were graphically checked to satisfy the correct inequality and thus were stable with respect to axisymmetric perturbations.

We now describe our numerical approach to obtain approximations to solutions to \eqref{eq:C3=0_fOmega} and \eqref{eq:C1C2=0_fOmega}. We first note that we can rescale the functions in consideration using the boundary condition \eqref{eq:BC_omega2},
\begin{equation*}
  f(x)
  =
  C_\omega\tilde{f}(x)
  \qquad\text{ and }\qquad
  \Omega(x)
  =
  C_\omega\tilde{\Omega}(x),
\end{equation*}
so that the boundary condition \eqref{eq:BC_omega2} becomes
\begin{equation}
\label{eq:BC_omega=1}
  \lim_{x\to1}\tilde{\Omega}(x)
  =
  1.
\end{equation}
Note that we can simply replace $f$ and $\Omega$ in \eqref{eq:C3=0_fOmega} and \eqref{eq:C1C2=0_fOmega} by $\tilde{f}$ and $\tilde{\Omega}$, since the scaling constants cancel out. We will thus work with \eqref{eq:C3=0_fOmega} and \eqref{eq:C1C2=0_fOmega} in their original form and only replace \eqref{eq:BC_omega2} with \eqref{eq:BC_omega=1}, keeping in mind that any solutions will correspond to the rescaled functions.

Disregarding the boundary condition on $\Omega$, it is clear that if the pair $(f,\Omega)$ solves \eqref{eq:C3=0_fOmega} and \eqref{eq:C1C2=0_fOmega}, then so does any pair $(\pm\tilde{c}f,\pm\tilde{c}\Omega)$ with $\tilde{c}\in\Real$. In our numerical approach we will seek solutions with $f>0$ and $\Omega>0$ in $(0,1)$. Note that if $f$ and $\Omega$ satisfy \eqref{eq:sol_f_omega}, then equation \eqref{eq:C3=0_fOmega} will be automatically satisfied. It is not difficult to check that the pair
\begin{equation}
\label{eq:f0Omega0}
  f_0(x)
  =
  2^{(1-b)/2}\left(x(1-x)\right)^{(2-b)/2}
  \qquad\text{ and }\qquad
  \Omega_0(x)
  =
  \left(\frac{2x}{1+x}\right)^{(1-b)/2}
\end{equation}
satisfies \eqref{eq:C3=0_fOmega} and the left-hand side of \eqref{eq:C1C2=0_fOmega} evaluates to a well-behaved expression
\begin{equation*}
   2^{1-b}(2-b)(1-b)\frac{2+x}{1+x}\left(x(1-x)\right)^{1-b}
\end{equation*}
that vanishes at both endpoints for all $0<b<1$ and converges uniformly to $0$ as $b\to1$. The expressions in \eqref{eq:f0Omega0} can therefore serve as a basis for initial guesses in a numerical scheme. However, we observe that their derivatives behave singularly near the endpoints. To bypass this difficulty, we recall \eqref{eq:f_asymptotics} and define a function $\gamma(x)$ by
\begin{equation}
\label{eq:f_gamma}
  f(x)
  =
 \gamma(x) (1-x^2)^{(2-b)/2},
\end{equation}
so that, using \eqref{eq:sol_omega_f},
\begin{equation}
\label{eq:Omega_gamma}
  \Omega(x)
  =
  c(1-x^2)^{-(1-b)/2}f(x)^{(1-b)/(2-b)}
  =
  c\gamma(x)^{(1-b)/(2-b)}
  \quad\text{ for some }c>0.
\end{equation}
We can then substitute \eqref{eq:f_gamma} and \eqref{eq:Omega_gamma} into \eqref{eq:C1C2=0_fOmega}. Since $\gamma$ is expected to have an infinite slope at $x=0$, we also reformulate the newly obtained version of \eqref{eq:C1C2=0_fOmega} in terms of the square of $\gamma$,
\begin{equation*}
  p(x)
  =
  \gamma^2(x),
\end{equation*}
and, after factoring out and discarding some positive terms, get the equation
\begin{equation}
\label{eq:C1C2=0p}
  \begin{split}
    p^2
    \biggr[
      (1-x^2)
      \big(
        (1-x^2)p'''
        &-
        2(4-b)xp''
      \big)
      -
      2(2+b-3(2-b)x^2)p'
    \biggr]
    +
    2c^2(1-b)p^{(3-2b)/(2-b)}p'\\
    &+
    \frac{1-b}{2-b}(1-x^2)p'
    \biggr[
      (1-x^2)(p')^2
      -
      2p
      \left(
        (1-x^2)p''
        -
        (2-b)xp'
      \right)
    \biggr]
    =
    0.
  \end{split}
\end{equation}
The boundary conditions arising immediately from those for $f$ are $\gamma(0)=0$ and, in view of the asymptotic behavior of $f$ given in \eqref{eq:f_asymptotics}, $\gamma$ having a finite limit as $x\to1$. Since the solutions can be rescaled as discussed above, we can assume $\gamma(x)\to1$ as $x\to1$. We approximate the solution to \eqref{eq:C1C2=0p} by discretizing it using a uniform mesh with stepsize $h$ and solving the discretized system by Newton's method, using \eqref{eq:f0Omega0} and \eqref{eq:f_gamma} to assemble an initial guess for $p(x)$. The results presented in this section correspond to $h=10^{-3}$.

While it is not clear to us whether solutions to \eqref{eq:C1C2=0p} exist for any combination of the constants $b$ and $c$, we have found that for a given value of $b$, increasing the value of $c$ eventually creates instability in the numerical code, suggesting a potential restriction on (a combination of) these values. Note from, e.g., \eqref{eq:f_gamma} and \eqref{eq:Omega_gamma} that the constant $c$ can be viewed as a scaling constant between $\Omega$ and $f$ (or, more generally, between the azimuthal component of the flow and the non-azimuthal ones); a large value of $c$ corresponds to a relatively large azimuthal component of the velocity with respect to the other two components, while a small value of $c$ corresponds to the azimuthal component being relatively small. In other words, increasing $c$ can be viewed as increasing swirl in the flow. Notice the similarity to changing $C_1$ in the case $b=1$ above.

In Fig.~\ref{fig:FOmegaG_c=0.25}, we present computed solutions for $c=0.25$ and $b=0.1,\dots,0.9$ with increments of $0.1$. Notice that the results demonstrate continuous dependence on the parameter $b$, and the solution \eqref{eq:sol_b=1} for $b=1$ can be viewed as their limit as $b\to1$. This solution, with constants $C_\omega=1$ and $C_1=4\sqrt{2}$, is plotted in Fig.~\ref{fig:FOmegaG_c=0.25} in red for comparison.
\begin{figure}
  \centering
  \includegraphics[width=0.32\textwidth]{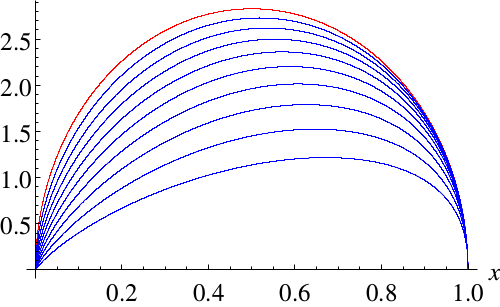}
  \hfill
  \includegraphics[width=0.32\textwidth]{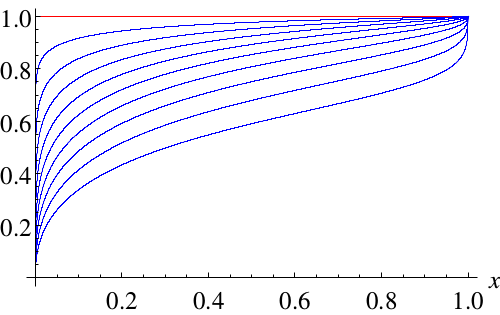}
  \hfill
  \includegraphics[width=0.32\textwidth]{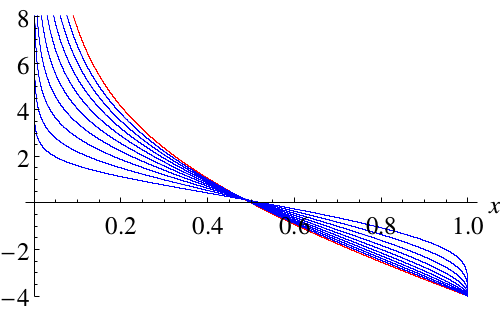}
  \caption{Graphs of $F(x)$ (left), $\Omega(x)$ (middle), and $G(x)$ (right) as numerical solutions obtained from \eqref{eq:C1C2=0p} with $c=0.25$ for $b=0.1,0.2,\dots,0.9$ (in blue). The functions $F$ and $\Omega$ are increasing with increasing $b$ at all $x\in(0,1)$, while the magnitude of $G$ increases in most of the interval $(0,1)$ as $b$ increases. The red plot corresponds to the solution \eqref{eq:sol_b=1} with $C_\omega=1$ and $C_1=4\sqrt{2}$, which can be viewed as a limiting case as $b\to1$.}
  \label{fig:FOmegaG_c=0.25}
\end{figure}
We can observe that in most of the interval $(0,1)$ the magnitudes of $F$, $\Omega$, and $G$ decrease as $b$ decreases, although the results do not suggest that these functions would vanish if $b$ approached $0$. Since the expression $1/r^b$ also decreases with decreasing $b$ for $0<r<1$, we see that if $c$ is fixed, then the flow speed decreases with decreasing $b$ near the vortex axis. This is consistent with observations of Cai \cite{cai} and Wurman \cite{wurman00,wurman05} that larger values of $b$ correspond to stronger storms.

The azimuthal velocity exhibits an interesting feature for $0<b<1$. Notice in Fig.~\ref{fig:FOmegaG_c=0.25} that $\Omega(0)=0$ for every $0<b<1$, and therefore the azimuthal velocity vanishes at the ground. This behavior of $\Omega$ is not enforced by an {\it a priori} boundary condition, rather is it a consequence of the Euler equations and the boundary condition $F(0)=0$. It means that nontrivial solutions with $0<b<1$ exhibit purely radial inflow or outflow at the ground.

To see how the choice of $c$ affects the results, we also present results with a fixed value of $b$ and varying values of $c$ for which we were able to generate results. In Fig.~\ref{fig:FOmegaG_b=0.6}, we present results with $b=0.6$ and $c=0.1,\dots,1.0$ with increments of $0.1$. (The value $b=0.6$ is chosen since it corresponds to the midpoint of the interval $(-0.7,-0.5)$ found in \cite{wurman05}.) For comparison, we also display the graphs corresponding to $c=0.25$ in red; these same graphs are also shown in Fig.~\ref{fig:FOmegaG_c=0.25}.
\begin{figure}
  \centering
  \includegraphics[width=0.32\textwidth]{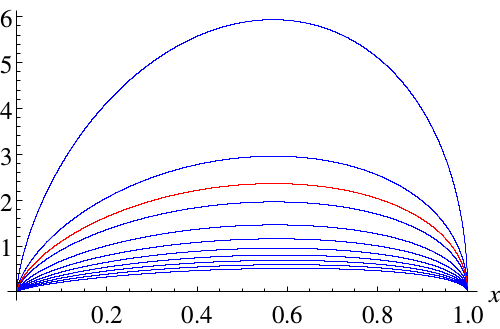}
  \hfill
  \includegraphics[width=0.32\textwidth]{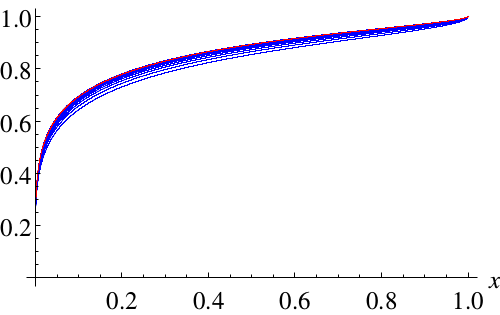}
  \hfill
  \includegraphics[width=0.32\textwidth]{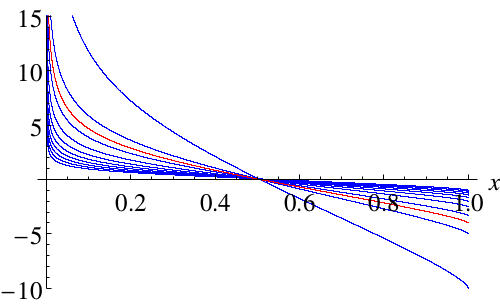}
  \caption{Graphs of $F(x)$ (left), $\Omega(x)$ (middle), and $G(x)$ (right) as numerical solutions obtained from \eqref{eq:C1C2=0p} with $b=0.6$ for $c=0.1,0.2,\dots,1.0$ (in blue). The functions $F$ and $\Omega$ are decreasing with increasing $c$ at all $x\in(0,1)$, while the magnitude of $G$ decreases in most of the interval $(0,1)$ as $c$ increases. The red plots correspond to the solutions with $c=0.25$, which are also shown in Fig.~\ref{fig:FOmegaG_c=0.25}.}
  \label{fig:FOmegaG_b=0.6}
\end{figure}
We see that as $c$ increases, the magnitudes of $F$, $\Omega$, and $G$ decrease in most of the interval $(0,1)$, although the change in $\Omega$ is fairly small. This behavior is in agreement with the meaning of the constant $c$ discussed earlier, i.e., that $c$ reflects the relative importance of the azimuthal component of the velocity with respect to the other two components.

To compare the flows, speeds, and pressure fields, we present in Fig.~\ref{fig:b=0.8_0.2} the analogs of Fig.~\ref{fig:b=1} for the cases $c=0.25$ and $b=0.8$ and $0.2$. The plots corresponding to the intermediate values of $b$ showed continuous dependence on the parameter $b$ and consequently we do not display them. We observe that while the streamlines remain relatively the same for various values of $b$, the speeds of the flow and the pressure fields exhibit discernible changes. The speeds are significantly larger in the plot with the larger value of $b$ (top row), and the figures also suggest a wider funnel for the larger $b$. Both of these observations are consistent with larger values of $b$ being associated with more violent storms.

\begin{figure}
  \centering
  \includegraphics[width=0.32\textwidth]{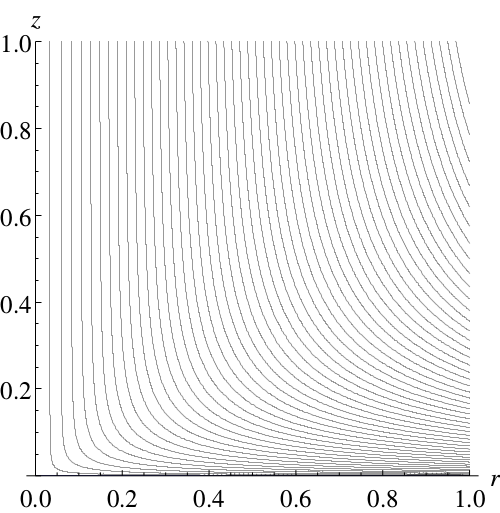}
  \hfill
  \includegraphics[width=0.32\textwidth]{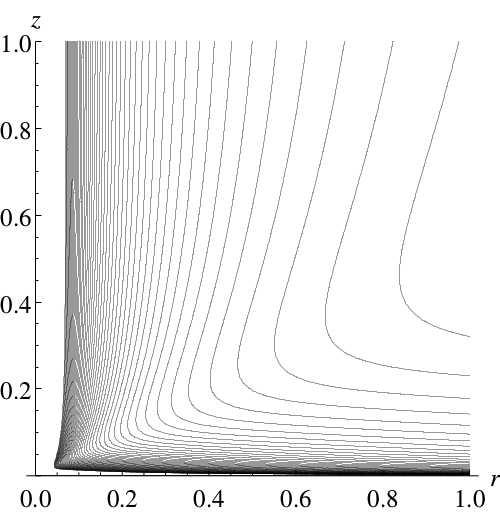}
  \hfill
  \includegraphics[width=0.32\textwidth]{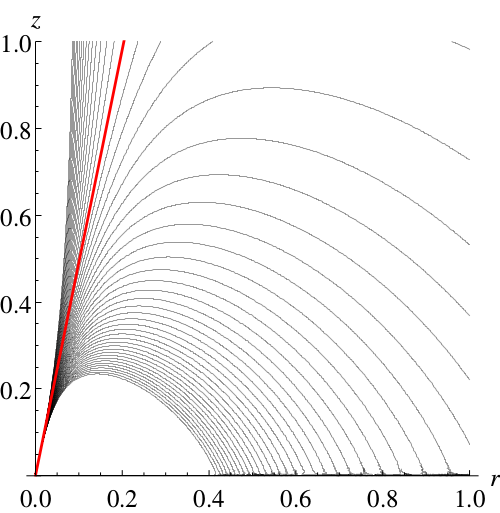}
  \vskip\baselineskip
  \includegraphics[width=0.32\textwidth]{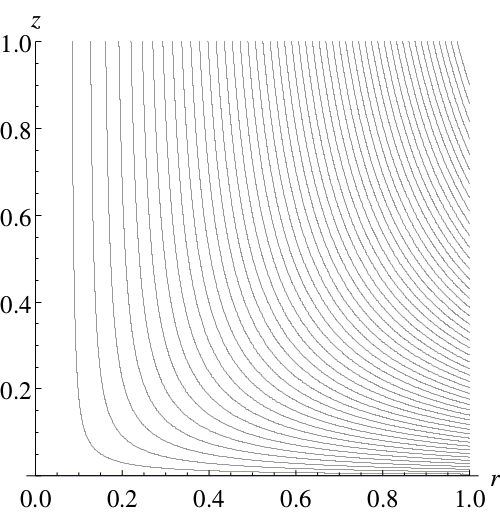}
  \hfill
  \includegraphics[width=0.32\textwidth]{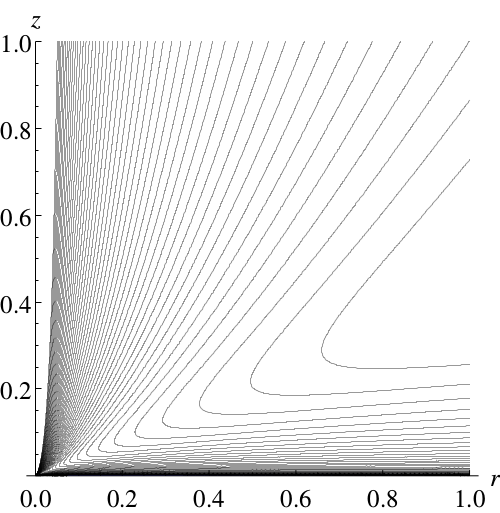}
  \hfill
  \includegraphics[width=0.32\textwidth]{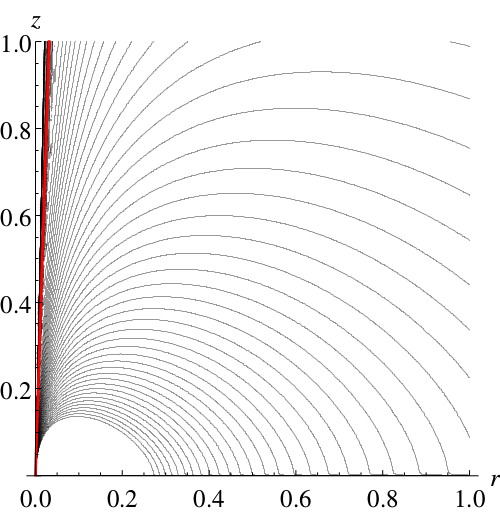}
  \caption{Plots of the streamlines (left), the corresponding isocurves for speed (middle), and isocurves for pressure (right) for the numerically computed solution with $c=0.25$, $b=0.8$ (top row) and $c=0.25$, $b=0.2$ (bottom row). The values for speed range from $3.20$ to $34.56$ with a step of $0.64$ (top), and from $1.7$ to $6.6$ with a step of $0.1$ (bottom), both increasing towards the vortex axis. The values for pressure range from $-24.0$ to $54.4$ with a step of $1.6$ (top), and from $1.82$ to $14.56$ with a step of $0.26$ (bottom), with lowest values near the vortex axis. (The contours near the vortex axis for $b=0.2$ are significantly affected by numerical errors in the contour plot routine.) The straight red lines correspond to the level sets $p(R,\alpha)=0$.}
  \label{fig:b=0.8_0.2}
\end{figure}

Our numerical results in this section were obtained under the assumption that $\Omega$ does not change sign in $(0,1)$. We do not yet know whether our equations \eqref{eq:C3=0_fOmega} and \eqref{eq:C1C2=0_fOmega} admit solutions that change sign in the interval $(0,1)$. Such cases would be interesting to study, since, unlike in \cite{serrin}, equation \eqref{eq:C3=0_fOmega} suggests that if $f$ changes sign, so would $\Omega$ and vice versa. This could give rise to very interesting types of flows.

\section{Discussion, Conclusions, and Implications}
\label{sec:conclusions}
In this work, we focused on finding solutions of the form \eqref{eq:velFGO} to the Navier--Stokes and Euler equations in the upper half-space. We were motivated by the 1972 work of J.~Serrin \cite{serrin}, in which he studies the viscous case with $b=1$, and by later studies \cite{cai,wurman00,wurman05}, in which it is suggested that the velocity may decay with different powers of the radial distance from the vortex axis than $1$. We found that no solutions of the form \eqref{eq:velFGO} exist in the case of constant nonzero viscosity when $b\ne1$, primarily due to the boundary conditions at the ground, i.e., the plane bounding our half-space.

The situation is different in the case with zero viscosity and Euler equations. In this case the boundary conditions are relaxed since slip is allowed. Assuming that the azimuthal velocity behaves like $C_\omega/r^b$ with $C_\omega\ne0$ near the vortex axis, we were able to show that the trivial solution $\Omega\equiv C_\omega$ and $F=G\equiv0$ works for any $b>0$, although it is stable with respect to axisymmetric perturbations only if $0<b\le1$.

Nontrivial solutions are harder to find. We showed that if $b\ge2$, no nontrivial solutions exist. We also showed that any potential solutions for $1<b<2$ would be unstable with respect to axisymmetric perturbations. The case with $b=1$ was fully analytically resolved and its solution is given in \eqref{eq:sol_b=1}. We discussed its characteristics and showed that the downdraft ($C_1>0$) solution in \eqref{eq:sol_b=1} can be viewed as a limit as viscosity goes to zero of the downdraft solutions found in \cite{serrin}.

The case with $0<b<1$ proved analytically difficult, and we only presented some numerical results indicating the existence of solutions that have similar characteristics to those with $b=1$. This case is most interesting, as it allows the coefficient $b$ to fall into the ranges discussed by Cai and Wurman et al.~\cite{cai,wurman00,wurman05}. We were able to numerically find solutions with $F$ and $\Omega$ that do not change sign, which corresponds to either an updraft or a downdraft solution. It would be interesting to see whether solutions with sign changes are possible. The numerically found solutions exhibit continuous dependence on the parameter $b$ and tend to the solution \eqref{eq:sol_b=1} as $b\to1$ as demonstrated in Fig.~\ref{fig:FOmegaG_c=0.25}. We have shown how the value of $b$ affects the intensity of the modeled vortex.

As in Serrin's model, our model exhibits a singularity near the vortex axis; in particular, the updraft/downdraft and azimuthal speeds tend to infinity, although at a rate of $1/r^b$, rather than $1/r$. As we discussed briefly in the introduction, updraft wind speeds may exceed the speed of sound, which may occur during the process of a vortex breakdown, illustrated in Fig.~\ref{fig:breakdown}. During this process, a single-cell flow bifurcates into a double-cell flow, and then further bifurcates into a flow with multiple vortices. The portion of the vortex near the axis where the horizontal flow turns into vertical (updraft case) is called the corner flow region. If we assume that this part of the flow with large vertical updraft speeds is quasi-steady, our and Serrin's models can be viewed as describing the lower portion of the flow. Additionally, extremely intense vortices with large updraft speeds can develop inside larger tornadoes as evidenced by the formation of ``suction spot'' paths in crop fields, paths less than $1$ meter in diameter where corn crops have been ripped out of the ground by the roots \cite{fujita81}. Such vortices could potentially be described using our and Serrin's models as well. We believe that current radar research in \cite{cai,wurman00,wurman05}, giving the power-law drop for velocity where $b\ne1$ ($2$ for vorticity) in tornadoes but varies, justifies our approach to modify Serrin's model. We have also provided numerical evidence that some of our solutions are viscosity solutions and hence remove the singular behavior near the ground, where velocity should tend to $0$.

Finally, we remark that in the case of a turbulent motion the functions of the form \eqref{eq:velFGO} correspond to the mean velocity, and thus $\nu$ could play a role of eddy viscosity to maintain the delicate balance between the mean and turbulent components of the flow. In recent studies, testing eddy viscosity assumptions with direct numerical simulations showed varied success \cite{chenkatzmeneveau05}, but has remained an important tool for understanding the connection between the scale of the model and dissipation of energy \cite{guermondodenprudhomme04,nolan05}. Following \cite{schlichting}, Serrin suggests an experimentally motivated relationship $\nu\approx\sigma\|{\bf v}\|r$, where $\sigma$ is a small dimensionless constant \cite{serrin}. Taking into account \eqref{eq:velfgo}, we thus obtain $\nu\approx\tau(x)R^{1-b}$, with the simplest case being $\tau\equiv\text{const}$. Such an assumption on viscosity then leads to a modification of equations \eqref{eq:ns333n} and \eqref{eq:ns12n} that will be investigated in the future.

\section{Appendix}

\subsection{Navier--Stokes equations in spherical coordinates and incompressibility}
The three components of the Navier--Stokes equations \eqref{eq:nss} expressed in spherical coordinates and in terms of the velocity components \eqref{eq:velRAT} have the form
\begin{equation}
\label{eq:ns1}
\begin{split}
  v_R\frac{\partial v_R}{\partial R}
  +
  \frac{v_\alpha}{R}\frac{\partial v_R}{\partial\alpha}
  +
  \frac{v_\theta}{R\sin\alpha}\frac{\partial v_R}{\partial\theta}
  &-
  \frac{v^2_\alpha+v^2_\theta}{R}
  =\\
  -\frac{\partial p}{\partial R}
  +
  \frac{\nu}{R^2}
  &\Biggr[
  \frac{\partial }{\partial R}\left(R^2\frac{\partial v_R}{\partial R}\right)
  +
  \frac{1}{\sin\alpha}\frac{\partial}{\partial\alpha}
    \left(\sin\alpha\frac{\partial v_R}{\partial\alpha}\right)\\
  &\quad+
  \frac{1}{\sin^2\alpha}\frac{\partial^2v_R}{\partial\theta^2}
  -
  2\left(
     v_R+\frac{\partial v_\alpha}{\partial\alpha}+v_\alpha\cot\alpha
     +
     \frac{1}{\sin\alpha}\frac{\partial v_\theta}{\partial\theta}
   \right)
  \Biggr],
\end{split}
\end{equation}
\begin{equation}
\label{eq:ns2}
\begin{split}
  v_R\frac{\partial v_\alpha}{\partial R}
  +
  \frac{v_\alpha}{R}\frac{\partial v_\alpha}{\partial\alpha}
  +
  \frac{v_\theta}{R\sin\alpha}\frac{\partial v_\alpha}{\partial\theta}
  &+
  \frac{v_Rv_\alpha-v^2_\theta\cot\alpha}{R}
  =\\
  -\frac{1}{R}\frac{\partial p}{\partial\alpha}
  &+
  \frac{\nu}{R^2}
  \Biggr[
  \frac{\partial}{\partial R}\left(R^2\frac{\partial v_\alpha}{\partial R}\right)
  +
  \frac{1}{\sin\alpha}\frac{\partial}{\partial\alpha}
  \left(\sin\alpha\frac{\partial v_\alpha}{\partial\alpha}\right)\\
  &\qquad\qquad+
  \frac{1}{\sin^2\alpha}\frac{\partial^2v_\alpha}{\partial\theta^2}
  +
  2\frac{\partial v_R}{\partial\alpha}
  -
  \frac{1}{\sin^2\alpha}
  \left(v_\alpha+2\cos\alpha\frac{\partial v_\theta}{\partial\theta}\right)
  \Biggr],
\end{split}
\end{equation}
\begin{equation}
\label{eq:ns3}
\begin{split}
  v_R\frac{\partial v_\theta}{\partial R}
  +
  \frac{v_\alpha}{R}\frac{\partial v_\theta}{\partial\alpha}
  +
  \frac{v_\theta}{R\sin\alpha}\frac{\partial v_\theta}{\partial\theta}
  &+
  \frac{v_Rv_\theta+v_\alpha v_\theta\cot\alpha}{R}
  =\\
  -\frac{1}{R\sin\alpha}\frac{\partial p}{\partial\theta}
  +
  \frac{\nu}{R^2}
  &\Biggr[
  \frac{\partial}{\partial R}\left(R^2\frac{\partial v_\theta}{\partial R}\right)
  +
  \frac{1}{\sin\alpha}\frac{\partial}{\partial\alpha}
  \left(\sin\alpha\frac{\partial v_\theta}{\partial\alpha}\right)\\
  &\quad+
  \frac{1}{\sin^2\alpha}\frac{\partial^2v_\theta}{\partial\theta^2}
  +
  \frac{1}{\sin^2\alpha}
  \left(2\sin\alpha\frac{\partial v_R}{\partial\theta}+2\cos\alpha\frac{\partial v_\alpha}{\partial\theta}-v_\theta\right)
  \Biggr].
\end{split}
\end{equation}
Similarly, the continuity equation \eqref{eq:ces} has the form
\begin{equation}
\label{eq:ces1}
  \frac{1}{R^2}\frac{\partial}{\partial R}\left(R^2v_R\right)
  +
  \frac{1}{R\sin\alpha}
  \left[
    \frac{\partial}{\partial\alpha}\left(v_\alpha\sin\alpha\right)
    +
    \frac{\partial v_\theta}{\partial\theta}
  \right]
  =
  0.
\end{equation}

\subsection{Expressions $C_i$ and $D_i$ for general $b>0$; the continuity equation}
Substituting variables \eqref{eq:velFGO} into the Navier--Stokes equations \eqref{eq:ns1}--\eqref{eq:ns3} and comparing with the forms in \eqref{eq:ns11}--\eqref{eq:ns33}, we obtain
the expressions for $C_i(\alpha)$ and $D_i(\alpha)$ (we omit the argument $x=\cos\alpha$ in the functions $F$, $G$, $\Omega$, $f$, $g$, and $\omega$, and, for example, we write $G'$ instead of $G'(\cos\alpha)$)
\begin{align*}
  C_1(\alpha)
  &=
  -\frac{F^2+b\,G^2+\Omega^2+b\,FG\cot\alpha+FG'\sin\alpha}{(\sin\alpha)^{2b}},\\
  C_2(\alpha)
  &=
  \frac{-(b\,F^2+\Omega^2)\cot\alpha+(1-b)FG-FF'\sin\alpha}{(\sin\alpha)^{2b}},\\
  C_3(\alpha)
  &=
  \frac{(1-b)G\Omega+F\left[(1-b)\Omega\cot\alpha-\Omega'\sin\alpha\right]}{(\sin\alpha)^{2b}},
\end{align*}
and
\begin{align*}
  D_1(\alpha)
  &=
  \frac{
    G''\sin^2\alpha
    -
    2(1-b)G'\cos\alpha
    -
    (1-b^2-\cos{2\alpha})G\csc^2\alpha
    -
    2(1-b)F\cot\alpha
    +
    2F'\sin\alpha
  }{(\sin\alpha)^b},\\
  D_2(\alpha)
  &=
  \frac{
    F''\sin^2\alpha
    -
    2(1-b)F'\cos\alpha
    -
    (1-b^2)F\csc^2\alpha
    -
    2b\,G\cot\alpha
    -
    2G'\sin\alpha
  }{(\sin\alpha)^b},\\
  D_3(\alpha)
  &=
  \frac{
    \Omega''\sin^2\alpha
    -
    2(1-b)\Omega'\cos\alpha
    -
    (1-b^2)\Omega\csc^2\alpha
  }{(\sin\alpha)^b}.
\end{align*}
In terms of $x$, and the functions $F$, $G$, $\Omega$, and $f$, $g$, $\omega$, these expressions can be written as
\begin{equation*}
  \begin{split}
    C_1(x)
    &=
    (1-x^2)^{-b}\left[-F^2-b\,G^2-\Omega^2-F\left(\sqrt{1-x^2}\,G'+b\dfrac{x}{\sqrt{1-x^2}}\,G\right)\right]\\
    &=
    (1-x^2)^{-1}
    \left[
      -f^2-b\,g^2-\omega^2-f\left(\sqrt{1-x^2}\,g'+\dfrac{x}{\sqrt{1-x^2}}\,g\right)
    \right],\\
  \end{split}
\end{equation*}
\begin{equation*}
  \begin{split}
    C_2(x)
    &=
    (1-x^2)^{-b}
    \left[
      -\dfrac{x}{\sqrt{1-x^2}}(b\,F^2+\Omega^2)
      -
      F\left(\sqrt{1-x^2}\,F'-(1-b)G\right)
    \right]\\
    &=
    (1-x^2)^{-1}
    \left[
      -\dfrac{x}{\sqrt{1-x^2}}(f^2+\omega^2)
      -
      f\left(\sqrt{1-x^2}\,f'-(1-b)g\right)
    \right],\\
  \end{split}
\end{equation*}
\begin{equation}
\label{eq:C3_FGO}
  \begin{split}
    C_3(x)
    &=
    (1-x^2)^{-b}\left[(1-b)G\Omega-F\left(\sqrt{1-x^2}\,\Omega'-(1-b)\dfrac{x}{\sqrt{1-x^2}}\,\Omega\right)\right]\\
    &=
    (1-x^2)^{-1}
    \left[
      (1-b)g\omega-\sqrt{1-x^2}\,f\omega'
    \right],
  \end{split}
\end{equation}
and
\begin{equation*}
  \begin{split}
    D_1(x)
    &=
    (1-x^2)^{-b/2}
    \left[
      (1-x^2)G''
      -
      2(1-b)x\,G'
      -
      \dfrac{2(1-x^2)-b^2}{1-x^2}\,G
      -
      2(1-b)\dfrac{x}{\sqrt{1-x^2}}\,F
      +
      2\sqrt{1-x^2}\,F'
    \right]\\
    &=
    (1-x^2)^{-b/2}
    \left[
      (1-x^2)g''
      +
      \left(\dfrac{1}{1-x^2}-(2-b)(1+b)\right)g
      +
      2\sqrt{1-x^2}\,f'
    \right],\\
  \end{split}
\end{equation*}
\begin{equation*}
  \begin{split}
    D_2(x)
    &=
    (1-x^2)^{-b/2}
    \left[
      (1-x^2)F''
      -
      2(1-b)x\,F'
      -
      \dfrac{1-b^2}{1-x^2}\,F
      -
      2b\dfrac{x}{\sqrt{1-x^2}}\,G
      -
      2\sqrt{1-x^2}\,G'
    \right]\\
    &=
    (1-x^2)^{-b/2}
    \left[
      (1-x^2)f''
      -
      b(1-b)f
      -
      2\dfrac{x}{\sqrt{1-x^2}}\,g
      -
      2\sqrt{1-x^2}\,g'
    \right],\\
  \end{split}
\end{equation*}
\begin{equation}
\label{eq:D3_FGO}
  \begin{split}
    D_3(x)
    &=
    (1-x^2)^{-b/2}
    \left[
      (1-x^2)\Omega''
      -
      2(1-b)x\,\Omega'
      -
      \dfrac{1-b^2}{1-x^2}\,\Omega
    \right]\\
    &=
    (1-x^2)^{-b/2}
    \left[
      (1-x^2)\omega''
      -
      b(1-b)\,\omega
    \right].
  \end{split}
\end{equation}
The expressions $\dot{C}_1+2b\,C_2$ and $\dot{D}_1+(1+b)D_2$ are then
\begin{equation}
\label{eq:C1C2_FGO}
  \begin{split}
    \dot{C}_1+2b\,C_2
    &=
    (1-x^2)^{-b}
    \Big[
      2b\dfrac{x}{\sqrt{1-x^2}}\left((1-b)F^2+b\,G^2\right)
      +
      2\sqrt{1-x^2}
      \left(
        (1-b)FF'
        +
        b\,GG'
        +
        \Omega\Omega'
      \right)\\
      &\qquad+
      (1-x^2)
      \left(
        F'G'
        +
        FG''
      \right)
      +
      b\dfrac{3-2b-2(1-2b)x^2}{1-x^2}FG
      +
      b\,x\,F'G
      -
      (1-3b)x\,FG'
    \Big],\\
    =
    (1-&x^2)^{-1}
    \Big[
      \dfrac{2x}{\sqrt{1-x^2}}\left((1-b)f^2+b\,g^2+(1-b)\omega^2\right)
      +
      2\sqrt{1-x^2}
      \left(
        (1-b)ff'
        +
        b\,gg'
        +
        \omega\omega'
      \right)\\
      &\qquad\qquad+
      (1-x^2)
      \left(
        f'g'
        +
        fg''
      \right)
      +
      \dfrac{1+2x^2+2b(1-b)(1-x^2)}{1-x^2}fg
      +
      x\,f'g
      +
      2x\,fg'
    \Big],
  \end{split}
\end{equation}
and
\begin{equation*}
  \begin{split}
    \dot{D}_1+(1+b)\,D_2
    &=
    (1-x^2)^{-(2+b)/2}
    \Biggr[
      (1-b)(1-b^2-2b(1-x^2))F
      -
      b^2(4+b-2x^2)\frac{x}{\sqrt{1-x^2}}\,G\\
      &\qquad\qquad\qquad+
      2(1-b)^2x(1-x^2)F'
      +
      (2-4b-b^2-2(1-3b+b^2)x^2)\sqrt{1-x^2}\,G'\\
      &\qquad\qquad\qquad-
      (1-x^2)^2
      \left[
        (1-b)F''
        -
        (4-3b)\frac{x}{\sqrt{1-x^2}}\,G''
        +
        \sqrt{1-x^2}\,G'''
      \right]
    \Biggr]\\
    &=
    (1-x^2)^{-3/2}
    \Biggr[
      -b(1-b^2)(1-x^2)f
      +
      (-3-b(1+b)(1-x^2))\frac{x}{\sqrt{1-x^2}}\,g\\
      &\qquad\qquad\qquad\qquad-
      (1+b(1+b)(1-x^2))\sqrt{1-x^2}\,g'\\
      &\qquad\qquad\qquad\qquad-
      (1-x^2)^2
      \left[
        (1-b)f''
        -
        \frac{x}{\sqrt{1-x^2}}\,g''
        +
        \sqrt{1-x^2}\,g'''
      \right]
    \Biggr].
  \end{split}
\end{equation*}

Finally, substituting \eqref{eq:velFGO} and \eqref{eq:velfgo} into the continuity equation \eqref{eq:ces1}, we get
\begin{equation}
\label{eq:ces23}
  \begin{split}
    R^{-(1+b)}(1-x^2)^{-b/2}
    \left[
      (2-b)G
      -
      \left(
        \sqrt{1-x^2}\,F'
        -
        (1-b)\dfrac{x}{\sqrt{1-x^2}}\,F
      \right)
    \right]
    =
    0,\\
    R^{-(1+b)}(1-x^2)^{-1/2}
    \left[
      (2-b)g
      -
      \sqrt{1-x^2}\,f'
    \right]
    =
    0.
  \end{split}
\end{equation}

\subsection{Expressions $C_i$ and $D_i$ for $b\ne2$} In this case, we can use equations \eqref{eq:G-F} and eliminate $G(x)$ and $g(x)$ from the expressions in the previous section. We have
\begin{equation}
\label{eq:C3_FGO_n2}
  \begin{split}
    C_3
    &=
    (1-x^2)^{1/2-b}
    \left[
      \frac{1-b}{2-b}\,\Omega(x)
      \left(
        F'(x)+\frac{x}{1-x^2}\,F(x)
      \right)
      -
      F(x)\Omega'(x)
    \right]\\
    &=
    (1-x^2)^{-1/2}
    \left[
      \frac{1-b}{2-b}f'(x)\omega(x)-f(x)\omega'(x)
    \right],
  \end{split}
\end{equation}
\begin{equation}
\label{eq:D3_FGO_n2}
  \begin{split}
    D_3
    &=
    (1-x^2)^{-1-b/2}
    \left[
      (1-x^2)\Omega''(x)
      -
      2(1-b)x(1-x^2)\Omega'(x)
      -
      (1-b^2)\Omega(x)
    \right]\\
    &=
    (1-x^2)^{-1/2}
    \left[
      (1-x^2)\omega''(x)-b(1-b)\omega(x)
    \right],
  \end{split}
\end{equation}
\begin{equation}
\label{eq:C1C2_FGO_n2}
  \begin{split}
    \dot{C}_1+2b\,C_2
    =
    \frac{(1-x^2)^{1/2-b}}{(2-b)}
    &\Biggr[
      (1-x^2)\left(\frac{2+b}{2-b}F'(x)F''(x)+F(x)F'''(x)\right)
      +
      2(2-b)\Omega(x)\Omega'(x)\\
      &\quad-
      2\,\frac{1-b}{2-b}
      \biggr[
        \frac{2x(1+bx^2)}{(1-x^2)^2}F^2(x)
        +
        \frac{b+(2+3b)x^2}{1-x^2}F(x)F'(x)\\
        &\qquad\qquad\qquad+
        (2+b)x(F'(x))^2
        +
        (4-b)xF(x)F''(x)
      \biggr]
    \Biggr]\\
    =
    \frac{(1-x^2)^{-1/2}}{2-b}
    &\biggr[
      (1-x^2)\left(\frac{2+b}{2-b}f'(x)f''(x)+f(x)f'''(x)\right)
      +
      2(2-b)\omega(x)\omega'(x)\\
      &\quad+
      2(1-b)
      \left[
      (2-b)\frac{x}{1-x^2}\left(f^2(x)+\omega^2(x)\right)
      +
      2f(x)f'(x)
      \right]
    \biggr],
  \end{split}
\end{equation}
and
\begin{equation}
\label{eq:D1D2_FGO_n2}
  \begin{split}
    \dot{D}_1+(1+b)D_2
    =
    -\frac{(1-x^2)^{-2-b/2}}{2-b}
    &\biggr[
      (1-x^2)^4F^{(4)}(x)
      -
      4(2-b)x(1-x^2)^3F'''(x)\\
      &\quad-
      2(1-b)(3+b-2(3-b)x^2)(1-x^2)^2F''(x)\\
      &\quad-
      4b(1-b)(2+b-x^2)x(1-x^2)F'(x)\\
      &\quad-
      (1-b)(3-b(1-b-b^2-4(3+b)x^2+4x^4))F(x)
    \biggr]\\
    =
    -\frac{(1-x^2)^{-1/2}}{2-b}
    &\biggr[
      (1-x^2)^2f^{(4)}(x)
      -
      4x(1-x^2)f'''(x)\\
      &-
      2b(1-b)(1-x^2)f''(x)
      +
      b(1-b^2)(2-b)f(x)
    \biggr].
  \end{split}
\end{equation}

\subsection{Equations \eqref{eq:bne1_NS} for $b\ne2$}
When $b\ne2$, we can use expressions \eqref{eq:G-F} and substitute them into equations \eqref{eq:bne1_NS} to get, in terms of $F$ and $\Omega$,
\begin{equation}
\label{eq:C3=0_G-F}
  F\Omega'
  =
  \frac{1-b}{2-b}\left[F'+\frac{x}{1-x^2}\,F\right]\Omega,
\end{equation}
\begin{equation}
\label{eq:L3=0_G-F}
  (1-x^2)^2\Omega''
  -
  2(1-b)x(1-x^2)\Omega'
  -
  (1-b^2)\Omega
  =
  0,
\end{equation}
\begin{equation}
\label{eq:C1C2=0_G-F}
  \begin{split}
    2(2-b)
    &\Omega\Omega'
    +
    (1-x^2)\left[\frac{2+b}{2-b}\,F'F''+FF'''\right]\\
    &\quad=
    2\,\frac{1-b}{2-b}
   \left[
      2\frac{x(1+bx^2)}{(1-x^2)^2}\,F^2
      +
      \frac{b+(2+3b)x^2}{1-x^2}\,FF'
      +
      (2+b)x(F')^2
      +
      (4-b)xF(x)F''
    \right],
  \end{split}
\end{equation}
\begin{equation}
\label{eq:L1L2=0_G-F}
  \begin{split}
    (1-x^2)^4&F^{(4)}
    -
    4(2-b)x(1-x^2)^3F'''
    -
    (1-b)\Big[
    2(3+b-2(3-b)x^2)(1-x^2)^2F''\\
    &\quad+
    4bx(2+b-x^2)(1-x^2)F'
    +
    (3-b+b^2+b^3+4b(3+b)x^2-4bx^4)F
    \Big]
    =
    0,
  \end{split}
\end{equation}
or, in terms of $f$ and $\omega$,
\begin{equation*}
  f(x)\omega'(x)
  =
  \frac{1-b}{2-b}f'(x)\omega(x),
\end{equation*}
\begin{equation*}
  (1-x^2)\omega''(x)
  -
  b(1-b)\omega(x)
  =
  0,
\end{equation*}
\begin{equation*}
  (1-x^2)
  \left[
    \frac{2+b}{2-b}\,f'f''
    +
    ff'''
  \right]
  +
  4(1-b)ff'
  +
  2(1-b)(2-b)
  \frac{x}{1-x^2}
  \left(
    f^2+\omega^2
  \right)
  +
  2(2-b)\omega\omega'
  =
  0,
\end{equation*}
\begin{equation*}
  (1-x^2)^2f^{(4)}
  -
  4x(1-x^2)f'''
  -
  2b(1-b)(1-x^2)f''
  +
  b(1-b^2)(2-b)f
  =
  0.
\end{equation*}

\bibliography{alt_powers}

\begin{thebibliography}{10}

\bibitem{bluestein07}
H.~B. Bluestein.
\newblock Advances in applications of the physics of fluids to severe weather
  systems.
\newblock {\em Rep. Prog. Phys.}, 70(8):1259--1323, 2007.

\bibitem{cai}
H.~Cai.
\newblock Comparison between tornadic and nontornadic mesocyclones using the
  vorticity (pseudovorticity) line technique.
\newblock {\em Mon. Wea. Rev.}, 133(9):2535--2551, 2005.

\bibitem{chenkatzmeneveau05}
J.~Chen, J.~Katz, and C.~Meneveau.
\newblock Implication of mismatch between stress and strain-rate in turbulence
  subjected to rapid straining and destraining on dynamic {LES} models.
\newblock {\em J. Fluids Eng.}, 127(5):840--850, 2005.

\bibitem{chorin}
A.~J. Chorin.
\newblock {\em Vorticity and turbulence}.
\newblock Springer-Verlag, New York, 1994.

\bibitem{davies-jones73}
R.~P. Davies-Jones.
\newblock The dependence of core radius on swirl ratio in a tornado simulator.
\newblock {\em J. Atmos. Sci.}, 30(7):1427--1430, 1973.

\bibitem{dipernamajda87}
R.~J. Di{P}erna and A.~J. Majda.
\newblock Oscillations and concentrations in weak solutions of the
  incompressible fluid equations.
\newblock {\em Commun. Math. Phys.}, 108(4):667--689, 1987.

\bibitem{dokken12}
D.~P. Dokken, K.~Scholz, M.~M. Shvartsman, P.~B\v{e}l\'{\i}k, C.~Potvin,
  B.~Dahl, and A.~McGovern.
\newblock Possible implications of a vortex gas model and self-similarity for
  tornadogenesis and maintenance.
\newblock Preprint online: \href{http://arxiv.org/abs/1403.0197}{\tt
  http://arxiv.org/abs/1403.0197}, 2014.

\bibitem{drazinreid}
P.~G. Drazin and W.~H. Reid.
\newblock {\em Hydrodynamic stability}.
\newblock Cambridge University Press, Cambridge, 2004.

\bibitem{fiedler94}
B.~H. Fiedler.
\newblock The thermodynamic speed limit and its violation in axisymmetric
  numerical simulations of tornado-like vortices.
\newblock {\em Atmos. Ocean}, 32(2):335--359, 1994.

\bibitem{fiedler96}
B.~H. Fiedler.
\newblock The sonic speed limit of tornadoes.
\newblock In {\em 18th Conference on Severe Local Storms}, pages 385--386.
  Amer. Meteor. Soc., 1996.

\bibitem{fiedlergarfield10}
B.~H. Fiedler and G.~S. Garfield.
\newblock Axisymmetric vortex simulations with various turbulence models.
\newblock {\em CFD Letters}, 2(3):112--122, 2010.

\bibitem{fiedlerrotunno86}
B.~H. Fiedler and R.~Rotunno.
\newblock A theory for the maximum windspeed in tornado-like vortices.
\newblock {\em J. Atmos. Sci.}, 43(21):2328--2440, 1986.

\bibitem{fujita81}
T.~T. Fujita.
\newblock Tornadoes and downbursts in the context of generalized planetary
  scales.
\newblock {\em J. Atmos. Sci.}, 38(8):1511--1534, 1981.

\bibitem{goldshtik60}
M.~A. Go\v{l}dshtik.
\newblock A paradoxical solution of the {N}avier--{S}tokes equations.
\newblock {\em J. Appl. Math. Mech.}, 24(4):913--929, 1960.
\newblock Translated from {\it Prikladnaya Mekhanika i Matematika},
  24(4):610--621, 1960.

\bibitem{goldshtikshtern88}
M.~A. Go\v{l}dshtik and V.~N. Shtern.
\newblock Conical lows of fluid with variable viscosity.
\newblock {\em Proc. R. Soc. Lond. A}, 419(1856):91--106, 1988.

\bibitem{guermondodenprudhomme04}
J.-L. Guermond, J.~T. Oden, and S.~Prudhomme.
\newblock Mathematical perspectives on large eddy simulation models for
  turbulent flows.
\newblock {\em J. Math. Fluid Mech.}, 6(2):194--248, 2004.

\bibitem{hamada07}
S.~Hamada.
\newblock Numerical solutions of {S}errin's equations by double exponential
  transformation.
\newblock {\em Publ. Res. Inst. Math. Sci.}, 43(3):795--817, 2007.

\bibitem{lewellens07}
D.~C. Lewellen and W.~S. Lewellen.
\newblock Near-surface intensification of tornado vortices.
\newblock {\em J. Atmos. Sci.}, 64(7):2176--2194, 2007.

\bibitem{lewellens07a}
D.~C. Lewellen and W.~S. Lewellen.
\newblock Near-surface vortex intensification through corner flow collapse.
\newblock {\em J. Atmos. Sci.}, 64(7):2195--2209, 2007.

\bibitem{lewellensxia00}
D.~C. Lewellen, W.~S. Lewellen, and J.~Xia.
\newblock The influence of a local swirl ratio on tornado intensification near
  the surface.
\newblock {\em J. Atmos. Sci.}, 57(4):527--544, 2000.

\bibitem{lewellenxialewellen02}
W.~S. Lewellen, J.~Xia, and D.~C. Lewellen.
\newblock Transonic velocities in tornadoes?
\newblock In {\em 21st Conference on Severe Local Storms}. Amer. Meteor. Soc.,
  2002.

\bibitem{long58}
R.~R. Long.
\newblock Vortex motion in a viscous fluid.
\newblock {\em J. Meteor.}, 15(1):108--112, 1958.

\bibitem{long61}
R.~R. Long.
\newblock A vortex in an infinite viscous fluid.
\newblock {\em J. Fluid Mech.}, 11(4):611--624, 1961.

\bibitem{madani03}
R.~Malek-Madani, J.~E. Coleman, and D.~R. Smith.
\newblock A numerical study of the swirling vortex.
\newblock Preprint online:
  \href{http://www.usna.edu/Users/math/rmm/SwirlingPaper.pdf}{\tt
  http://www.usna.edu/Users/math/rmm/SwirlingPaper.pdf}, 2003.

\bibitem{newton01}
P.~K. Newton.
\newblock {\em The {N}-vortex problem. Analytical Techniques}.
\newblock Springer-Verlag, New York, 2001.

\bibitem{nolan05}
D.~S. Nolan.
\newblock A new scaling for tornado-like vortices.
\newblock {\em J. Atmos. Sci.}, 62(7):2639--2645, 2005.

\bibitem{NIST}
F.~W.~J. Olver, D.~W. Lozier, R.~F. Boisvert, and C.~W. Clark, editors.
\newblock {\em NIST Handbook of Mathematical Functions}.
\newblock Cambridge University Press, Cambridge, 2010.

\bibitem{rotunno13}
R.~Rotunno.
\newblock The fluid dynamics of tornadoes.
\newblock {\em Annu. Rev. Fluid Mech.}, 45:59--84, 2013.

\bibitem{schlichting}
H.~Schlichting.
\newblock {\em Boundary-Layer Theory}.
\newblock McGraw-Hill, New York, 1960.

\bibitem{serrin}
J.~Serrin.
\newblock The swirling vortex.
\newblock {\em Phil. Trans. Roy. Soc. London, Series A, Math \& Phys. Sci.},
  271(1214):325--360, 1972.

\bibitem{shapiromarkowski99}
A.~Shapiro and P.~Markowski.
\newblock Dynamics of elevated vortices.
\newblock {\em J. Atmos. Sci.}, 56(9):1101--1122, 1999.

\bibitem{shternhussain99}
V.~Shtern and F.~Hussain.
\newblock Collapse, symmetry breaking, and hysteresis in swirling flows.
\newblock {\em Annu. Rev. Fluid Mech.}, 31:537--566, 1999.

\bibitem{wang91}
C.~Y. Wang.
\newblock Exact solutions of the steady-state {N}avier--{S}tokes equations.
\newblock {\em Annu. Rev. Fluid Mech.}, 23:159--177, 1991.

\bibitem{ward72}
N.~B. Ward.
\newblock The exploration of certain features of tornado dynamics using a
  laboratory model.
\newblock {\em J. Atmos. Sci.}, 29(6):1194--1204, 1972.

\bibitem{wu86}
J.~Z. Wu.
\newblock Conical turbulent swirling vortex with variable eddy viscosity.
\newblock {\em Proc. R. Soc. Lond. A}, 403(1825):235--268, 1986.

\bibitem{wurman05}
J.~Wurman and C.~R. Alexander.
\newblock The 30 {M}ay 1998 {S}pencer, {S}outh {D}akota, storm. {P}art {II}:
  {C}omparison of observed damage and radar-derived winds in the tornadoes.
\newblock {\em Mon. Wea. Rev.}, 133(1):97--119, 2005.

\bibitem{wurman00}
J.~Wurman and S.~Gill.
\newblock Finescale radar observations of the {D}immitt, {T}exas (2 {J}une
  1995), tornado.
\newblock {\em Mon. Wea. Rev.}, 128(7):2135--2164, 2000.

\bibitem{wurman13}
J.~Wurman, K.~Kosiba, and P.~Robinson.
\newblock In situ, {D}oppler radar, and video observations of the interior
  structure of a tornado and the wind--damage relationship.
\newblock {\em Bull. Amer. Met. Soc}, 94(6):835--846, 2013.

\bibitem{wurman96}
J.~Wurman, J.~M. Straka, and E.~N. Rasmussen.
\newblock Fine-scale {D}oppler radar observations of tornadoes.
\newblock {\em Science}, 272(5269):1774--1777, 1996.

\bibitem{xialewellens03}
J.~Xia, D.~C. Lewellen, and W.~S. Lewellen.
\newblock Influence of {M}ach number on tornado corner flow dynamics.
\newblock {\em J. Atmos. Sci.}, 60(22):2820--2825, 2003.

\bibitem{yih82}
C.-S. Yih, F.~Wu, A.~K. Garg, and S.~Leibovich.
\newblock Conical vortices: A class of exact solutions of the
  {N}avier--{S}tokes equations.
\newblock {\em Phys. Fluids}, 25(12):2147--2158, 1982.

\end{thebibliography}
\bibliographystyle{abbrv}
\end{document}